\documentclass[twocolumn]{aastex61}
\usepackage{longtable}
\accepted{to ApJ, \today}

\shorttitle{Testing Atomic Diffusion in M67}
\shortauthors{Souto et al.}

\begin{document}

\title{Chemical Abundances of Main-Sequence, Turnoff, Subgiant and red giant Stars from APOGEE spectra II: Atomic Diffusion in M67 Stars}

\correspondingauthor{Diogo Souto}
\email{souto@on.br, diogodusouto@gmail.com}

\author[0000-0002-7883-5425]{Diogo Souto}
\affiliation{Observat\'orio Nacional, Rua General Jos\'e Cristino, 77, 20921-400 S\~ao Crist\'ov\~ao, Rio de Janeiro, RJ, Brazil}

\author{C. Allende Prieto}
\affiliation{Instituto de Astrof\'isica de Canarias, E-38205 La Laguna, Tenerife, Spain}
\affiliation{Departamento de Astrof\'isica, Universidad de La Laguna, E-38206 La Laguna, Tenerife, Spain}

\author{Katia Cunha}
\affiliation{Observat\'orio Nacional, Rua General Jos\'e Cristino, 77, 20921-400 S\~ao Crist\'ov\~ao, Rio de Janeiro, RJ, Brazil}
\affiliation{Steward Observatory, University of Arizona, 933 North Cherry Avenue, Tucson, AZ 85721-0065, USA}

\author{Marc Pinsonneault}
\affiliation{Department of Astronomy, The Ohio State University, Columbus, OH 43210, USA}

\author{Verne V. Smith}
\affiliation{National Optical Astronomy Observatory, 950 North Cherry Avenue, Tucson, AZ 85719, USA}

\author{R. Garcia-Dias}
\affiliation{Instituto de Astrof\'isica de Canarias, E-38205 La Laguna, Tenerife, Spain}
\affiliation{Departamento de Astrof\'isica, Universidad de La Laguna, E-38206 La Laguna, Tenerife, Spain}

\author{Jo Bovy}
\affiliation{Department of Astronomy and Astrophysics, University of Toronto, 50 St. George Street, Toronto, ON, M5S 3H4, Canada}
\affiliation{Dunlap Institute for Astronomy and Astrophysics, University of Toronto, ON M5S 3H4, Canada}

\author{D. A. Garc\'ia-Hern\'andez}
\affiliation{Instituto de Astrof\'isica de Canarias, E-38205 La Laguna, Tenerife, Spain}
\affiliation{Departamento de Astrof\'isica, Universidad de La Laguna, E-38206 La Laguna, Tenerife, Spain}

\author{Jon Holtzman}
\affiliation{New Mexico State University, Las Cruces, NM 88003, USA}

\author{J. A. Johnson}
\affiliation{Department of Astronomy, The Ohio State University, Columbus, OH 43210, USA}

\author[0000-0002-4912-8609]{Henrik J\"onsson}
\affiliation{Lund Observatory, Department of Astronomy and Theoretical Physics, Lund University, Box 43, SE-221 00 Lund, Sweden}

\author{Steve R. Majewski}
\affiliation{Department of Astronomy, University of Virginia, Charlottesville, VA 22904-4325, USA}

\author{Matthew Shetrone}
\affiliation{University of Texas at Austin, McDonald Observatory, USA}

\author{Jennifer Sobeck}
\affiliation{Department of Astronomy, University of Virginia, Charlottesville, VA 22904-4325, USA}

\author{Olga Zamora}
\affiliation{Instituto de Astrof\'isica de Canarias, E-38205 La Laguna, Tenerife, Spain}
\affiliation{Departamento de Astrof\'isica, Universidad de La Laguna, E-38206 La Laguna, Tenerife, Spain}

\author{Kaike Pan}
\affiliation{Apache Point Observatory and New Mexico State University, P.O. Box 59, Sunspot, NM, 88349-0059, USA}

\author{Christian Nitschelm}
\affiliation{Centro de Astronom{\'i}a (CITEVA), Universidad de Antofagasta, Avenida Angamos 601, Antofagasta 1270300, Chile}

\begin{abstract}
Chemical abundances for 15 elements (C, N, O, Na, Mg, Al, Si, K, Ca, Ti, V, Cr, Mn, Fe, and Ni) are presented for 83 stellar members of the 4 Gyr old solar-metallicity open cluster M67.  The sample contains stars spanning a wide range of evolutionary phases, from G dwarfs to red clump stars.  
The abundances were derived from near-IR ($\lambda$1.5 -- 1.7$\mu$m) high-resolution spectra ($R$ = 22,500) from the SDSS-IV/APOGEE survey.
A 1-D LTE abundance analysis was carried out using the APOGEE synthetic spectral libraries, via chi-square minimization of the synthetic and observed spectra with the qASPCAP code. 
We found significant abundance differences ($\sim$0.05 -- 0.30 dex) between the M67 member stars as a function of the stellar mass (or position on the HR diagram), where the abundance patterns exhibit a general depletion (in [X/H]) in stars at the main-sequence turnoff.
The amount of the depletion is different for different elements.
We find that atomic diffusion models provide, in general, good agreement with the abundance trends for most chemical species, supporting recent studies indicating that measurable atomic diffusion operates in M67 stars.
\end{abstract}


\section{Introduction}

M67 (Messier 67; NGC 2886) is a well-studied open cluster, with an age and metallicity (4 Gyr and [Fe/H]=0.0, respectively) similar to those of the Sun. A number of studies have determined the distance to the cluster (\citealt{Yadav2008}), its age (\citealt{Yadav2008}, \citealt{Sarajedini2009}), photometric colors and reddening (\citealt{Taylor2007}, \citealt{Sarajedini2009}), as well as metallicity and individual chemical abundances (\citealt{Cohen1980}, \citealt{FoyProust1981}, \citealt{Tautvaisiene2000}, \citealt{Pancino2010}, \citealt{Jacobson2011}, \citealt{Onehag2014}, \citealt{Liu2016}, \citealt{Gao2018}, \citealt{BertelliMotta2018}, and \citealt{Souto2018}). M67 is a ``benchmark'' Galactic open cluster and an excellent laboratory in which to study poorly understood processes in stellar astrophysics, such as abundance variations in open clusters.

The chemical composition of a star is inherited from the interstellar matter from which it forms; however this composition changes over time due to internal stellar processes, such as gravitational settling or atomic diffusion. 
The approximation employed in the determination of abundances can also induce systematic errors in the inferred abundances creating an apparent lack of homogeneity.
Examples of such simplifications are the assumptions of hydrostatic equilibrium or local thermodynamical equilibrium (LTE).

Stellar clusters are useful astrophysical environments to study elemental abundance variations due to the reasonable assumption that stars in a cluster were born from the same molecular cloud at the same time. 
Several authors have studied the initial chemical homogeneity of open and globular clusters (\citealt{DeSilva2006,DeSilva2007}, \citealt{Reddy2012}, \citealt{Bovy2016}) and have, so far, not found any evidence of inhomogeneities in the initial stellar populations of open and globular clusters.

One well-known process that has been extensively observed in clusters is that as stars evolve into red giants, their surface carbon and nitrogen abundances are altered by the convectively-driven first dredge-up of material from the stellar interior that has been exposed to H-burning via the CN-cycle (\citealt{Lagarde2012}, \citealt{Bressan2012}, \citealt{Choi2016}). 
This process does not, however, explain the lack of uniformity in the elemental abundances of main-sequence and turnoff stars found in metal-poor globular clusters (\citealt{Korn2007}, \citealt{Lind2008}, \citealt{Nordlander2012}).
These variations are instead explained by atomic diffusion, a fundamental process predicted by theory (\citealt{Michaud2015}, references therein), and operating in all stars, which is often ignored in stellar evolution models and abundance studies. 
Atomic diffusion represents the physical process that involves the transport of material in the stellar atmosphere that is described by a diffusion equation, e.g., gravitational settling.
Atomic diffusion has a physical basis, with diffusion coefficients predicted by theory \cite{Chapman1917a,Chapman1917b}, \cite{Aller1960}, \cite{Michaud1976,Michaud1980}, \cite{Vauclair1978,Vauclair1982}, \cite{Michaud2004}.

Diffusion in stars having a solar age and metallicity, as is the case for members of M67, has been theoretically investigated by \cite{Michaud2004}, who analyzed 28 elements, finding that He, Li, Be, B, Mg, P, Ti, Fe, and Ni were those most affected by this mechanism. 
One of their conclusions was that atomic diffusion models can have a significant impact on the stellar ages derived from isochrones. 
More recently, theoretical calculations by \cite{Dotter2017} concluded that atomic diffusion also plays an important role in stars with a solar age and metallicity  (not only metal-poor stars), and found that the photospheric iron abundance in turnoff stars can be depleted by $\sim$0.12 dex compared to their initial surface abundance as a consequence of atomic diffusion processes. 
\cite{Dotter2017} noted that ignoring diffusion in models would cause an additional uncertainty of about 10\% in the stellar ages derived from isochrones.

Evidence for the occurrence of diffusion in M67 stars has been found previously by \cite{Onehag2014}, who studied a sample of fourteen stars belonging to M67, including main-sequence stars (6), turnoff (3), and the early subgiant branch (5), using high-resolution optical spectra from FLAMES/UVES on the VLT. \cite{Onehag2014} found abundance differences among the groups of 0.05--0.10 dex for Al, Ca, Cr, Mn, and Fe, with turnoff stars having lower abundances than subgiants.
\cite{Blanco-Cuaresma2015} compiled a sample of 42 stars in M67 (28 main-sequence and 14 red giants) using spectra from NARVAL, HARPS, and UVES. The authors observed that the abundances of Na, Mg, and Si show variations of up to 0.10--0.20 dex between dwarf and giant stars in the cluster.

\cite{Souto2018} (Paper I) studied a small sample of eight M67 stellar members spanning a range of evolutionary phases, including G-dwarfs (2), G-turnoff stars (2), G-subgiants (2), and red clump K-giants (2) using high-resolution spectra from the Apache Point Observatory Galactic Evolution Experiment (APOGEE; \citealt{Majewski2017}). 
They found abundance variations in fourteen elements across the HR diagram, confirming that most chemical species display changes in the range of 0.05--0.20 dex (Fe, Na, Mg, Al, Si, Ca, and Mn), with the lower abundances observed in turnoff stars, with M $\sim$1.2$M_{\odot}$. 
\cite{Souto2018} also showed that the abundance variations found in M67 stars compare very well with theoretical models of atomic diffusion for stars having the solar age and metallicity.
Also using APOGEE spectra, the study of \cite{BertrandeLis2016} found significantly more dispersion in [O/Fe] for M67 stars than for other clusters with similar metallicity but younger ages, such as NGC 6819 or NGC 2158. 
\cite{Bovy2016} and \cite{Price-Jones2018} found strong constraints on the chemical homogeneity in M67 red giant stars from APOGEE. The authors showed that M67 red giants are homogeneous based only on their stellar spectra, without the need of modeling the stellar atmosphere.  The uniformity within the red giant stars may indicate that changes in the stellar abundances across different evolutionary phases in the HR diagram for M67 might be related to physical processes operating within these stars.

The works of \cite{BertelliMotta2018} and \cite{Gao2018} have confirmed, using independent data, that atomic diffusion operates in M67 stars. 
Both works used high-resolution optical spectra; \cite{BertelliMotta2018} used UVES/FLAMES ($R$ $\sim$ 20,000--32,000) observations from the Gaia/ESO survey (\citealt{Gilmore2012}, \citealt{Randich2013}), reporting abundances of eleven elements in fifteen stars from the main-sequence, turnoff, and red giant branch. \cite{BertelliMotta2018}, using APOGEE data, find abundance variations of up to 0.20--0.30 dex for elements like Al, Mn, and Ni, where non-LTE effects are unlikely to explain the observed trends.
\cite{Gao2018} use spectra from the GALAH survey (\citealt{Galahsurvey2015}), with a resolving power of R$\sim$28,000, to report abundances for seven elements in 66 stars from the turnoff, subgiant, red giant, and red clump phases. \cite{Gao2018} conclude that deviations from non-LTE can explain some of the observed abundance trends as a function of the evolutionary stage, in particular for oxygen and sodium. However, for Al and Si, non-LTE does not explain the remaining trend, which the authors argue might be a consequence of diffusion processes in M67.

This work provides a complementary verification of the atomic diffusion mechanisms acting in M67 stars as reported by \cite{Souto2018}. We use APOGEE results obtained with the qASPCAP$\footnote{github.com/callendeprieto/}$ pipeline using a much larger stellar sample; qASPCAP is a simple IDL script that substitutes the entire ASPCAP (APOGEE Stellar Parameters and Chemical Abundances Pipeline, \citealt{GarciaPerez2016}) for boutique work.

APOGEE targeted M67 as one of its calibration clusters, observing about a hundred stellar members from the lower main-sequence, the turnoff, the subgiant branch, and the red giant branch. 
The M67 APOGEE sample is well-suited both to probe the limits on chemical homogeneity in the cluster members, as well as to search for signatures of atomic diffusion in the chemical abundances of a number of elements. APOGEE spectra are used here to derive detailed chemical abundances of fifteen elements: C, N, O, Na, Mg, Al, Si, K, Ca, Ti, V, Cr, Mn, Fe, and Ni. 

The paper is structured as follows: in Section 2 we describe the adopted sample, in Section 3 we report on the atmospheric parameters and the methodology employed to derive the individual abundances, in Section 4 we analyze the abundance trends, and in Section 5 we suggest possible explanations for them. In Section 6 we discuss the obtained results, summarizing in Section 7.

\section{The APOGEE Data on M67}

The APOGEE spectrographs are cryogenic multi-fiber near-infrared instruments covering the H-band between $\lambda$1.51 $\mu$m - $\lambda$1.69 $\mu$m, obtaining high-resolution (R=$\lambda$/$\Delta$$\lambda$  $\sim$ 22,500) spectra for 300 objects at a time (\citealt{Wilson2010}, \citealt{Gunn2006}). 
The spectrographs are currently mounted in both hemispheres on 2.5m telescopes at APO (Apache Point Observatory, New Mexico, USA) and at LCO (Las Campanas Observatory, La Serena, Chile). 
The M67 stellar spectra analyzed in this work were all obtained at APO, and reduced with the APOGEE pipeline, described in \cite{Nidever2015}.

APOGEE is part of the SDSS-III and SDSS-IV projects (\citealt{Eisenstein2011},\citealt{Blanton2017}) and M67 is one of the calibration clusters for the ASPCAP pipeline (\citealt{Zasowski2013}, \citealt{Meszaros2013}; \citealt{Holtzman2015}, \citealt{GarciaPerez2016}). 
APOGEE has observed a dedicated field in the direction of M67 (location ID 4162), obtaining spectra for 563 targets. 
The stars had multiple visits, generally more than three, to reach the required signal-to-noise ratio (SNR) of the combined spectra (higher than $\sim$100 per half a resolution element); this was achieved for stars brighter than $H$ $\leq$ 11.

\begin{figure*}
\begin{center}
\includegraphics[width=0.45\linewidth]{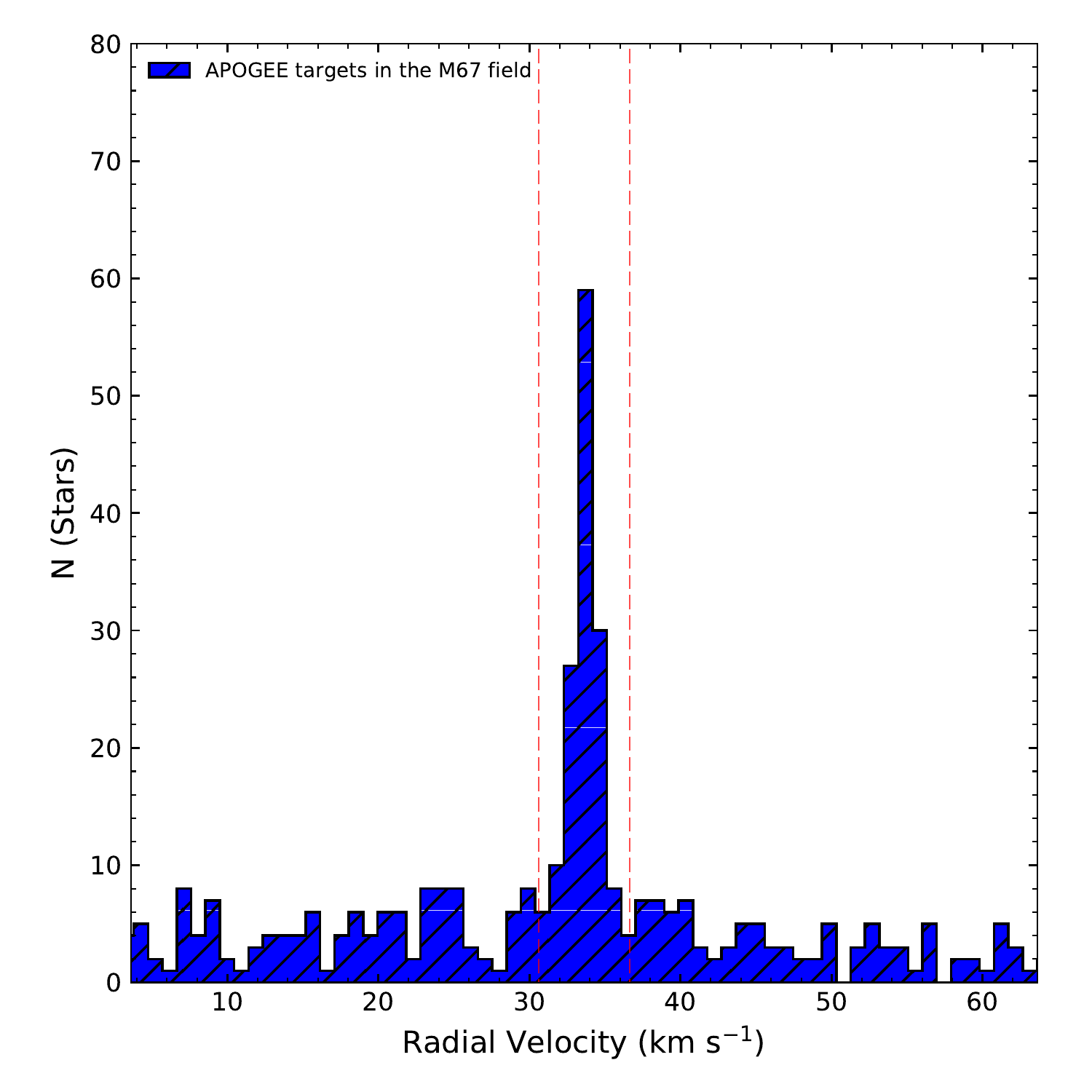}
\includegraphics[width=0.45\linewidth]{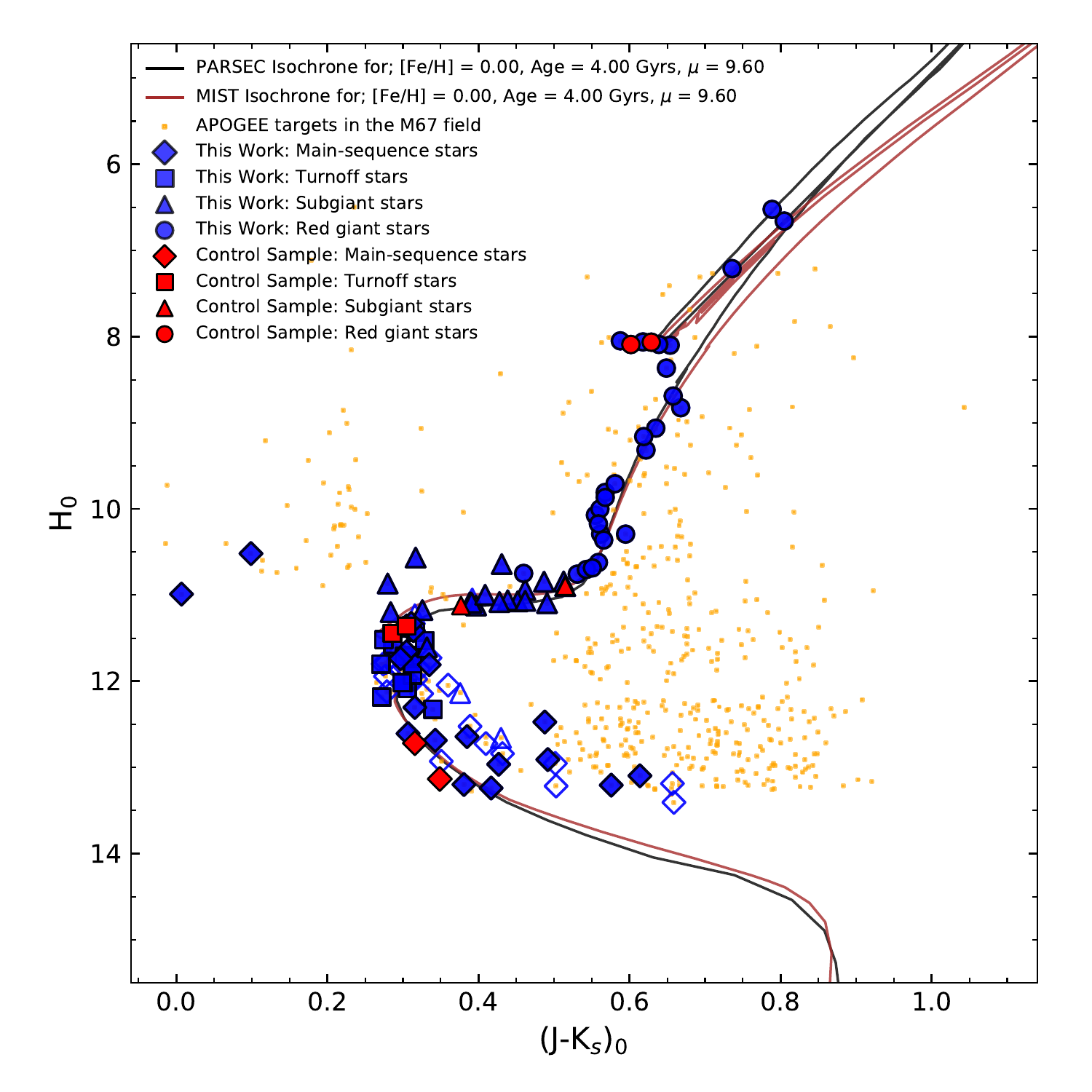}
\includegraphics[width=0.45\linewidth]{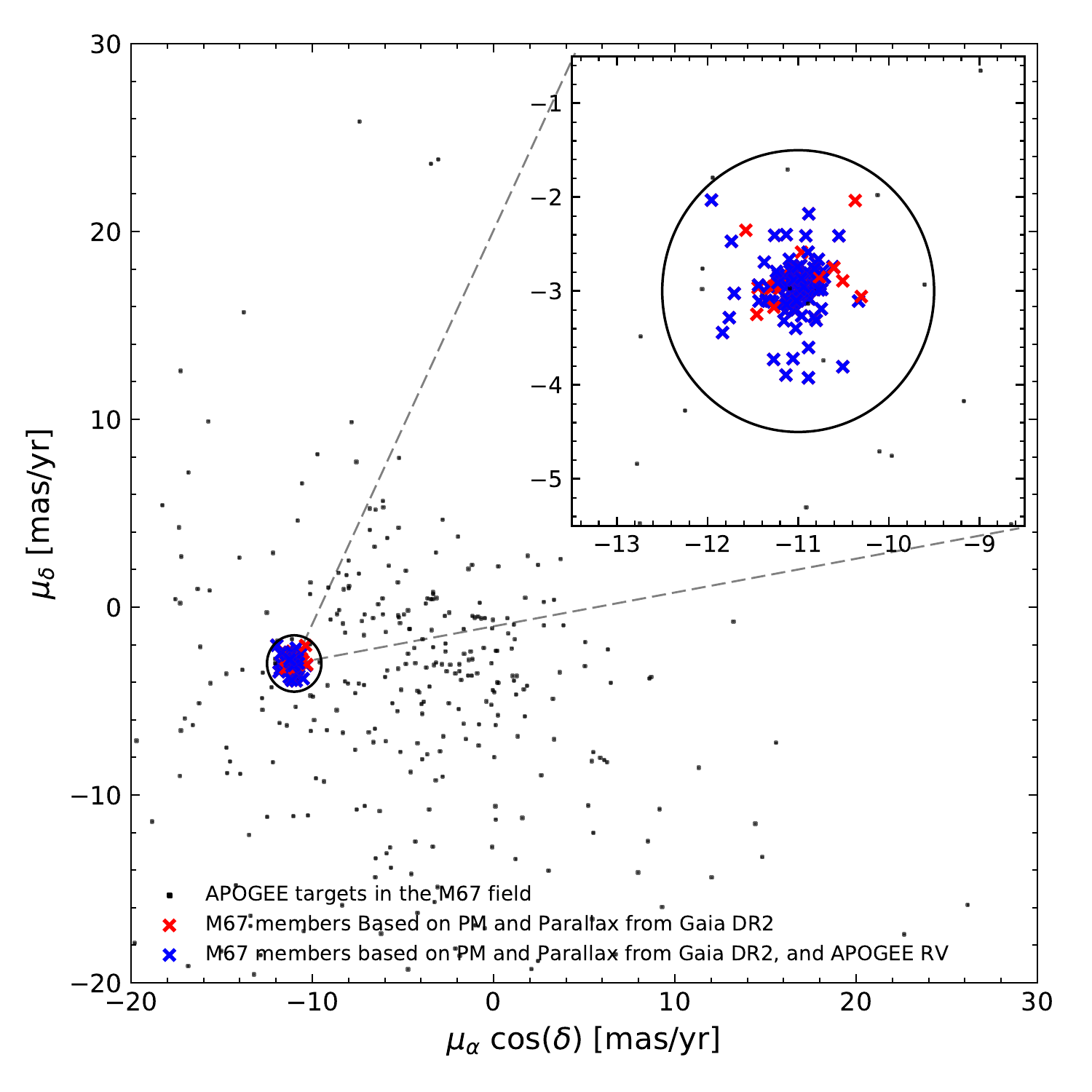}
\includegraphics[width=0.45\linewidth]{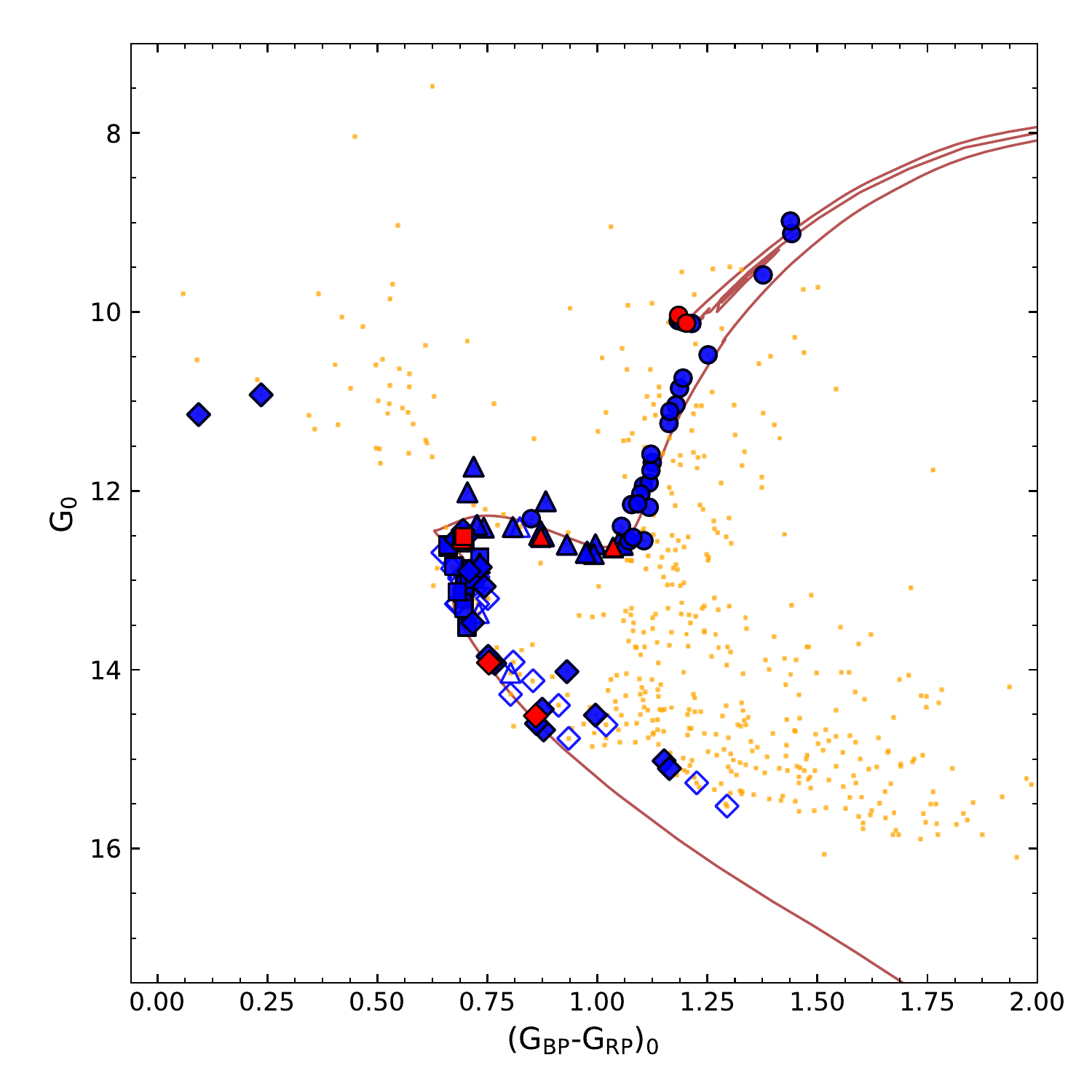}
\caption{Top left panel: Radial velocity distribution obtained from ASPCAP for all the targets observed in the APOGEE M67 field. 
Bottom left panel: vector-point diagram with the adopted stellar proper motions.
Top right panel: ($J$-$K_{s}$)$_{0}$ \textit{vs} $H_{0}$ diagram showing the sample of this work. We represent the main-sequence stars as blue diamonds, blue squares are the turnoff stars, the blue triangles for the subgiants, and the blue circles represent the red giant stars.
We also include the stellar sample of \cite{Souto2018} using the same symbol notation for the stellar classes; however, colored as red.
2MASS color-magnitude diagram of the APOGEE targets in the M67 field are shown as orange dots. 
Two isochrones for an age of 4 Gyr, (m-M)$_{0}$ = 9.60, and [Fe/H] = 0.00 from PARSEC (black line) and MIST (brown line) are also shown.
We left as open symbols the stars with SNR $<$ 100.
Bottom right panel: same as top right panel, expect the CMD using Gaia DR2 data for (G$_{\rm BP}$-G$_{\rm RP}$)$_{0}$ \textit{vs} $G$$_{0}$.}
\end{center}
\label{fig1_diagram}
\end{figure*}

To verify membership of the observed stars in the M67 APOGEE field, we adopt two approaches, one using membership studies from the literature and another using distances and proper motions from Gaia DR2 (\citealt{GaiaCollaborationDR2}).
We initially used the radial velocities ($RV$) measured by the APOGEE pipeline available in the 14th SDSS data release (DR14, \citealt{DR14}), following the proper motion and $RV$ membership criteria of \cite{Yadav2008} and \cite{Geller2015} as guidelines. 
\cite{Yadav2008} determined proper motions for 2462 stars using the Wide-Field-Imager from the MPG/ESO 2.2m telescope at La Silla, Chile, with a field of view of 34$\times$33 arcmin$^{2}$. The authors reported 434 stars having membership probabilities $\geq$ 90\%. Using the same data, \cite{Bellini2010} derived the cluster average proper motion to be 
$\mu_{\alpha}$ cos($\delta$) = -9.6 $\pm$ 1.1 mas yr$^{-1}$ and $\mu_{\delta}$ = -3.7 $\pm$ 0.8 mas yr$^{-1}$.
The radial velocity survey by \cite{Geller2015} used spectra obtained from various sources, including a total of 1278 stars in the vicinity of M67. 
\cite{Geller2015} reported 590 stars having membership probabilities $\geq$ 90\%, where the mean radial velocity of the sample is 33.64 km s$^{-1}$, with high internal precision (0.03 km s$^{-1}$).

Based on this information, an initial membership cut was performed, selecting from the targets observed in the M67 APOGEE field (563 stars), those within the radial velocity range 30.64 -- 36.64 km s$^{-1}$.
Figure \ref{fig1_diagram} (top-left panel) shows a histogram of the $RV$ distribution of all the stars in the field.
The peak of the $RV$ distribution compares well with the mean radial velocity for the cluster reported by \cite{Geller2015}, with 140 stars falling within the $RV$ limit (red dashed lines).
We then performed a cross-match between the stars within the limit in radial velocity and those stars reported by \cite{Yadav2008} and \cite{Geller2015} having membership probabilities $\geq$ 90 percent. A total of 119 stars satisfied these criteria.

We then adopted Gaia DR2 (\citealt{GaiaCollaborationDR2}) proper motions with distances from \cite{Bailer-Jones2018} to refine the sample.
From those 140 stars within the $RV$ limits, we find 109 within the ranges in distance and proper motion for M67. 
We accepted stars with distances in the range of 796.2 -- 992.0 pc, which corresponds to a distance modulus of 9.56--9.88, as reported in the literature for the cluster (\citealt{Yadav2008}, \citealt{Yakut2009}).
We then adopted the mean proper motions observed for the stars within the adopted distance limits, where $\mu_{\alpha}$cos($\delta$) = -11.02$\pm$0.07 mas/yr and $\mu_{\delta}$ = -2.97$\pm$0.05 mas/yr. 
We consider as members the stars within $\pm$ 1 mas/yr from those mean values.
Figure \ref{fig1_diagram} bottom left panel displays the proper motions for the sample.

We removed from the sample two hot stars (2M08512643 + 1143506 and 2M08513259+1148520) likely to be blue-stragglers.
In the final sample, we will only retain the stars with Gaia DR2 data, confirming the membership criteria based on distances and proper motions.
We searched for binary stars in our sample looking for RV variations in the multiple spectral visits, with none found. Also, we verify the lack of binary stars comparing our sample (44 stars in common) with the recent work of \cite{El-Badry2018},  where the authors detected more than 3000 binary stars in the APOGEE data.
To ensure the quality of the observed spectra, we keep only those having a signal to noise ratio SNR $\geq$ 100, resulting in a sample of 83 stars spanning the HR diagram, from the main-sequence to the red clump.
The threshold in SNR is intended to minimize the uncertainties in the parameters derived. As we are searching for small abundance variations across the HR diagram, we assemble the best possible sample.
We will include the results reported by \cite{Souto2018} as a control/comparison sample.
In Table 1 we present our sample, with the adopted radial velocity and SNR (from DR14), proper motions and distances (\citealt{GaiaCollaborationDR2}), membership probabilities computed by \cite{Geller2015} and the adopted magnitudes, $V$ (\citealt{URAT1}), and 2MASS infrared magnitudes $J$, $H$, and $K_{S}$ (\citealt{2MASS}). At the bottom of the table we also provide data for those stars with SNR $<$ 100.

In the top right and bottom right panels of Figure \ref{fig1_diagram} we display the color-magnitude diagram ($J$-$K_{S}$)$_{0}$ \textit{vs} $H_{0}$ and (G$_{\rm BP}$-G$_{\rm RP}$)$_{0}$ \textit{vs} $G$$_{0}$ for the studied sample using 2MASS and Gaia DR2 photometry, respectively. We show all 563 stars observed in the M67 field by the APOGEE survey with orange dots. Our sample stars are shown as filled symbols, and the ones with SNR $<$ 100 as empty symbols.
We note that four early G- and K- dwarfs show a small offset compared to the adopted isochrones presented in the CMD diagrams of Figure \ref{fig1_diagram}, which could indicate non-membership; however, we opt to use these stars as their RVs, proper motions, and distances suggest membership.	
The same symbol notation adopted by \cite{Souto2018} were used in this work, where diamonds correspond to main-sequence, squares to turnoff stars, triangles for subgiants, and the circles represent the red giant stars, in blue for this work and red for \cite{Souto2018}.

\begin{figure*}
\begin{center}
\includegraphics[width=0.8\linewidth]{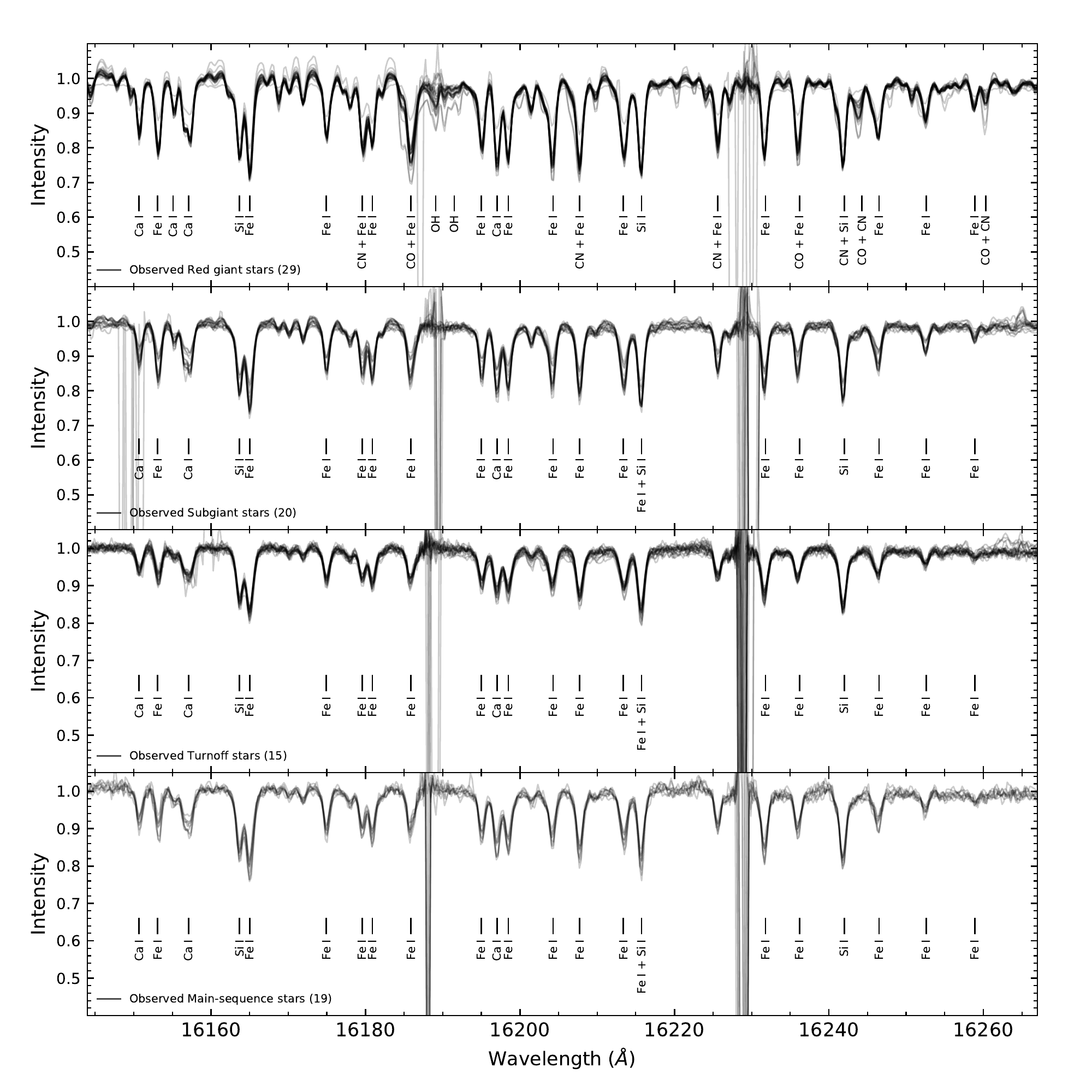}
\caption{A portion of the APOGEE observed spectra for the stellar sample. From top to bottom we shown the spectra of red giant, subgiant, turnoff, and main-sequence stars.}
\end{center}
\label{Fig_Spectra}
\end{figure*}

In Figure \ref{Fig_Spectra}, we display a portion of the observed APOGEE spectra between 16150--16260 \AA{} for the sample stars. 
From top to bottom, we plot the spectra of the red giant stars followed by the subgiant, turnoff, and main-sequence stars.
The individual stellar spectra are very similar within a class, with rms differences at any given wavelength of about $\sigma$ = 0.01.
The largest star-to-star differences in Figure \ref{Fig_Spectra} are associated with CO, CN, and OH lines in the red giant spectra, suggestive of the changes produced by H-burning in the stellar interior brought to the surface by the first dredge-up, as discussed in Section 5. Fe I and Ca I show the largest spread among G type stars.

\section{Stellar Parameters and Chemical Abundances}

In this paper, we need to determine abundances in different classes of stars (dwarfs to red giants) homogeneously and precisely. One important factor in such analysis is the determination of the stellar parameters. It is known that the raw log $g$ values derived using the ASPCAP pipeline contain systematic offsets for dwarfs (being systematically low) as well as red giant stars (being systematically high). 

Figure \ref{fig_HRdiag} shows the effective temperature and surface gravity diagrams for our sample. The left panel shows the DR14 raw ASPCAP $T_{\rm eff}$ and log $g$ results. It is clear that the log $g$ values derived by ASPCAP do not match the isochrones from \cite{Bressan2012} and \cite{Choi2016} (presented in the Figure). Using such log $g$ values in the analysis would introduce systematic uncertainties in the derived abundances. In the next section we discuss the determination of the log $g$'s and adopted $T_{\rm eff}$ values in this study.

\subsection{Effective Temperatures}

We adopted the effective temperatures derived from ASPCAP DR14. We used the purely spectroscopic raw $T_{\rm eff}$ values from ASPCAP (given in the FPARAM array in DR14).
For a comparison, we also determined photometric temperatures adopting the calibration of \cite{GonzalezHernandez2009} and using five different colors, $B$-$V$, $V$-$J$, $V$-$H$, $V$-$K_{s}$, and $J$-$K_{s}$, with an adopted cluster reddening of E(B-V) = 0.041 mag (\citealt{Sarajedini2009}) and a metallicity of [Fe/H] = 0.00.
\cite{GonzalezHernandez2009} provide photometric calibrations for red giant and dwarf stars; 
we adopted the coefficients for giants for those stars with log $g$ $<$ 4.00 dex and for dwarfs for those stars with higher gravities.
Good agreement between the photometric and the adopted raw ASPCAP $T_{\rm eff}$ scales is obtained, where $\langle$$\delta$($T_{\rm eff}$(ASPCAP - GHB)$\rangle$ = -25 $\pm$ 106 K.
The effective temperatures obtained from the ASPCAP pipeline have an internal precision of $\pm$ 50K (\citealt{Holtzman2015}, \citealt{GarciaPerez2016})

\subsection{Surface Gravities}

We determined surface gravities from the fundamental Equation 1, where the adopted $T_{\rm eff's}$ are from the raw ASPCAP DR14 values, with stellar masses and bolometric magnitudes obtained from interpolation in the MIST isochrones (\citealt{Choi2016}; [Fe/H] = 0.00; age = 4.00 Gyr; E(B-V)=0.041; distance modulus ($\mu$) = 9.60). 
The adopted solar values are: log $g_{\odot}$ = 4.438 dex, $T_{\rm eff, \odot}$ = 5772 K and $M_{bol,\odot}$ = 4.75, following the IAU recommendations in \cite{IAU_solar}.

\begin{center}
\begin{equation}
\log{g} = \log_{10}{g_{\odot}} + \log_{10}\left(\frac{M_{\star}}{M_{\odot}}\right) +
4\log_{10}\left(\frac{T_{\star}}{T_{\odot}}\right) + 0.4(M_{bol,\star} - M_{bol,\odot}),
\end{equation}\\
\end{center}

We adopted the surface gravities derived from equation 1 in the abundance analysis in this study.
The uncertainties in the determined surface gravities are similar to the ones reported in \cite{Souto2018}, where $\sigma$ = $\pm$ 0.10 dex.
The comparison between the derived log $g$'s in this work with those from ASPCAP confirm the log $g$ offset, where we obtain $\langle$$\delta$(log $g$(Physical - ASPCAP)$\rangle$ = -0.18 $\pm$ 0.16 dex for red giants, -0.16 $\pm$ 0.11 dex for subgiants, -0.19 $\pm$ 0.07 dex for turnoff and 0.17 $\pm$ 0.13 dex for the main-sequence stars.

\begin{figure*}
\begin{center}
\includegraphics[width=0.49\linewidth]{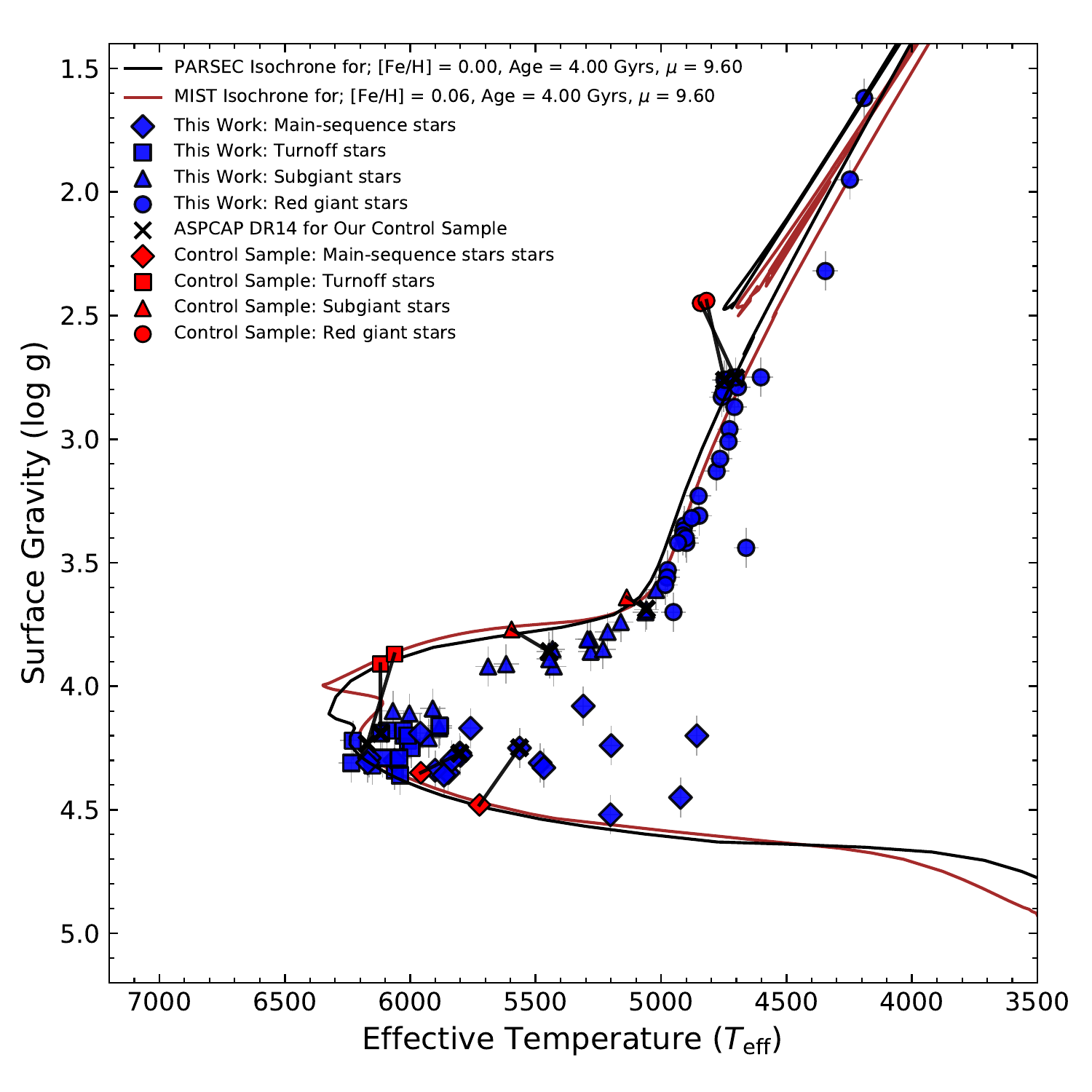}
\includegraphics[width=0.49\linewidth]{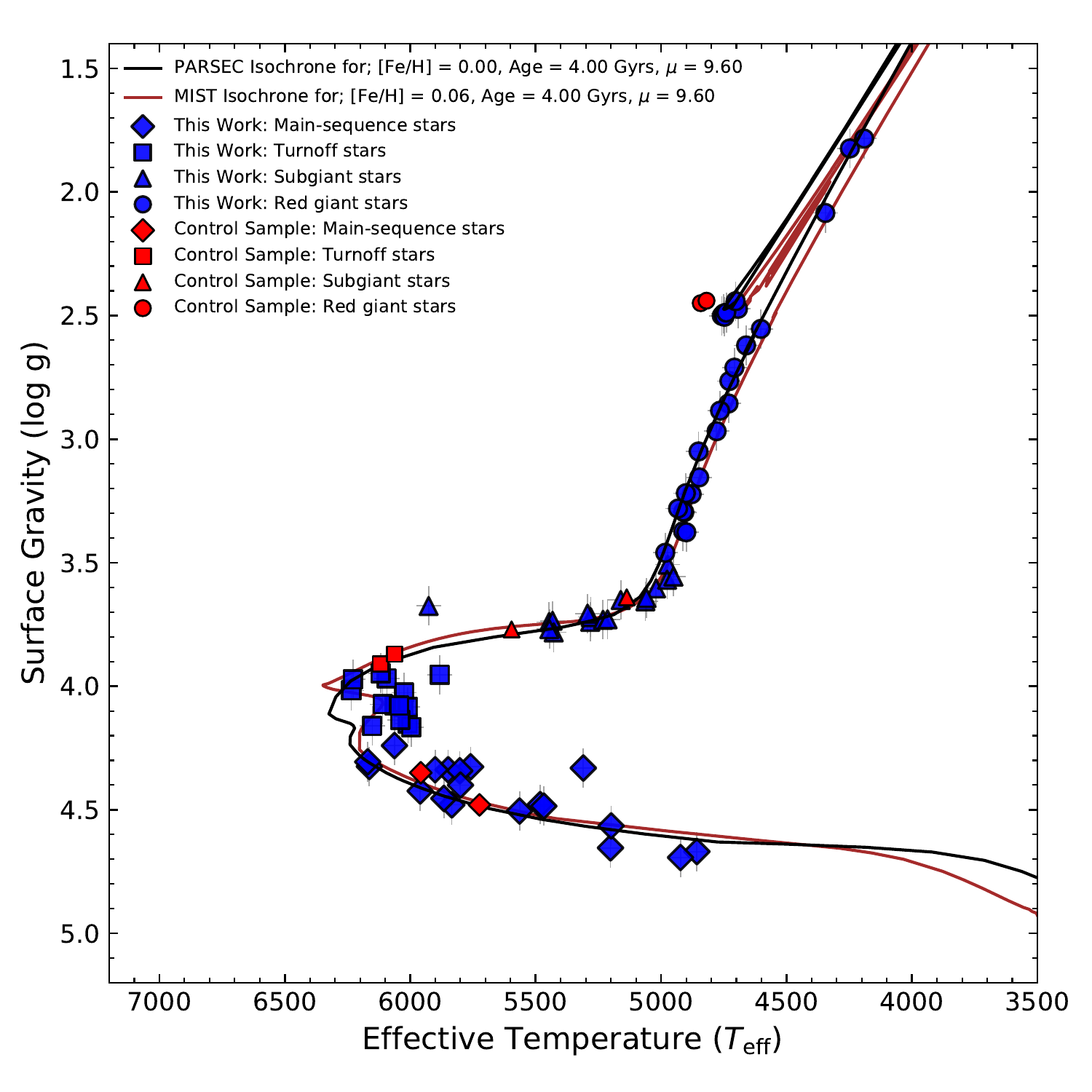}
\caption{
Left panel: $T_{\rm eff}$ -- log $g$ diagram showing the APOGEE DR14 raw ASPCAP results for the M67 members. Note the mismatch with the isochrones due to systematic uncertainties in the log $g$ values derived by ASPCAP.
Right panel: $T_{\rm eff}$ -- log $g$ diagram showing the stellar parameters adopted in this study.
The $T_{\rm eff's}$ are the same raw values from ASPCAP DR14 shown in the left panel but the surface gravities were derived from fundamental relations.
The symbol notation is the same as Figure \ref{fig1_diagram}.
}
\end{center}
\label{fig_HRdiag}
\end{figure*}

Figure \ref{fig_HRdiag} (right panel) shows the $T_{\rm eff}$ -- log $g$ values adopted in this study. 
The effective temperatures for the studied stars are well spread in the HR diagram, with effective temperatures ranging between 4200 and 6250 K. The surface gravity values for the studied stars span a range in log $g$ = 1.78 to 4.71.

\subsection{Individual Abundances Analysis}

In this work we derive individual abundances for fifteen elements: C, N, O, Na, Mg, Al, Si, K, Ca, Ti, V, Cr, Mn, Fe, and Ni. 
Individual abundances were determined with the qASPCAP code. The qASPCAP code basically corresponds to the ASPCAP pipeline, but for custom work, providing flexibility to change the analysis parameters.
The methodology in the analysis is the same as adopted in ASPCAP and the optimization is based on the FERRE code.

The procedure for determining individual abundances and microturbulent velocities with qASPCAP is similar to the one in ASPCAP.
The ASPCAP pipeline (described in detail in \citealt{GarciaPerez2016}) uses a grid of synthetic spectra (\citealt{Zamora2015}) computed with the turbospectrum code (\citealt{AlvarezPLez1998}, \citealt{Plez2012}) using KURUCZ model atmospheres (\citealt{CastelliKurucz2004}, \citealt{Meszaros2012}) and the APOGEE DR14 line list, which is an updated version of the one published in \cite{Shetrone2015}. 
The stellar parameters and chemical abundances are obtained by chi-square minimization with the FERRE code (\citealt{FERRE}) controlled by an IDL wrapper (the qASPCAP in this work).

In a first phase, seven parameters are determined through a 7-D optimization ($T_{\rm eff}$, log $g$, [M/H], [C/Fe], [N/Fe], [$\alpha$/Fe], and $\xi$) using the entire wavelength range of the APOGEE spectra. 
During the second phase, individual abundances are obtained by repeating the fitting in pre-determined windows that are sensitive to elemental abundances using the set of atmospheric parameters determined in the previous phase.
It is possible to determine individual abundances for more than 26 elements from the APOGEE spectra; see \cite{Holtzman2018}, \cite{Hasselquist2016} (for Nd) and \cite{Cunha2017} (for Ce).
In this work we adopt the same molecular and atomic lines as \cite{Souto2018} to derive individual abundances (see also \citealt{Smith2013} and \citealt{Souto2016}).
Even though \cite{Souto2018} reported Na and Cr abundances for main-sequence and turnoff stars, we opt in this work to not present these abundances (for these stellar classes) as the comparisons between the observed/synthesis were not satisfactory due to the weakness of the Na I and Cr I lines.

All M67 targets studied here have similar vsin($i$), between 0 $\leq$ vsin($i$) $\leq$ 7 km s$^{-1}$. In fact, the threshold to detect the star's vsin($i$) from APOGEE spectra is $\sim$ 7--8 km s$^{-1}$.
The effect of macroturbulence on the line profiles is similar to that of stellar rotation and, as an approximation, qASPCAP treats rotation and macroturbulence as a single Gaussian profile.

The stellar parameters adopted in this work are shown in Table 2, with individual abundances presented in Table 3.
The uncertainties in the derived abundances adopted in this work are the same as the ones reported in Table 4 of \cite{Souto2018}.
We note that using ASPCAP calibrated abundances, the average $\langle$$\delta$ A(El)$\rangle$ between the results derived in this work minus ASPCAP is smaller 0.10 dex for all elements.

\section{Results}

The individual abundances reported in this work display an elevated scatter (standard deviation of the mean), in particular for nitrogen ($\sim$0.14 dex), aluminum ($\sim$0.16 dex) and the alpha-elements ($\sim$0.15 dex). The potassium abundances are the ones showing the smallest scatter, with $\sigma$ = 0.07 dex.
Such significant scatter in M67 stars was also noticed by \cite{BertrandeLis2016} studying [O/Fe] in M67 stars with APOGEE and comparing it with the spread in other clusters.
However, when we analyze the stars by class (main-sequence, turnoff, subgiant, red giant), the scatter in the derived elemental abundance is drastically reduced to 0.03--0.04 dex for most of the elements.
As our sample covers a wide range in surface gravity,  1.78 $\leq$ log $g$ $\leq$ 4.71, it is possible that the observed scatter is the signature of a physical process modifying the stellar atmospheric abundances, such as atomic diffusion as proposed by \cite{Souto2018}. 
In the following sections, we discuss in detail the abundance trends as a function of the stellar parameters.

\subsection{Abundance Variations Across The H-R Diagram in M67 Stars:}

We split our sample into four different classes based on the stars' evolutionary stage. 
We selected as main-sequence stars those with log $g$ $\geq$ 4.20; turnoff stars those with surface gravity between 3.90 $<$ log $g$ $<$ 4.20; subgiants those having 3.60 $\leq$ log $g$ $\leq$ 3.90; and red giant stars those with log $g$ $<$ 3.60. 
(We note that the cut in surface gravity is similar to the one in color and magnitude, as can be seen in the right panel of Figure \ref{fig1_diagram}.)

Probing the level of homogeneity in open clusters is important to understand their formation and for evaluating the possibility of performing chemical tagging in stellar populations.
Chemical homogeneity in open clusters (as well as in globular clusters) is a critical assumption to understand changes in the abundances across evolutionary stages.
\cite{Bovy2016} and \cite{Price-Jones2018}, using APOGEE spectra, found tight constraints on the chemical homogeneity of M67 using a sample of red giant stars.
\cite{Bovy2016} analyzed 24 red giant stars in M67, finding one-dimensional sequences with a spread in the elemental initial cluster abundances lower than 0.03 dex (2 $\sigma$ of uncertainty) for all elements studied in this work.
It is worth noting that the \cite{Bovy2016} results were derived in a way that is insensitive to the effects of atomic diffusion, mixing, and other physical processes that may modify the stellar surface abundances.

One straightforward way to evaluate if samples of stars have similar abundances is to apply a Kolmogorov--Smirnov test (K-S test). 
The K-S test is usually invoked to find out if two samples are drawn from the same distribution.
We perform a study of chemical homogeneity of M67 stars using the derived abundances through a K-S test and we apply it to the same classes, e. g., red giants $\times$ red giants.
To be able to compare the derived abundances for the same classes with the K-S test, we randomly split each group into two samples and then we apply the K-S test. 
To ensure we do not choose a random split that favors homogeneity, for each group, we have run the test in one thousand random splits. 
This result shows that the abundances of each stellar class are indistinguishable, with the derived median $p$-value $>$ 0.50 for all elements in the four stellar classes. 
This is a complementary result to \cite{Bovy2016}, finding chemical homogeneity of M67 stars in the same evolutionary stage based on the stellar abundances derived in this work.

We also applied the K-S test using the derived abundances for the fifteen studied elements comparing stars in the different groups: G dwarf main-sequence (MS) $\times$ red giant; G dwarf (MS) $\times$ subgiant; G dwarf (MS) $\times$ turnoff; red giant $\times$ subgiant; red giant $\times$ turnoff; and subgiant $\times$ Turnoff stars.

\begin{figure*}
\begin{center}
\includegraphics[width=0.9\linewidth]{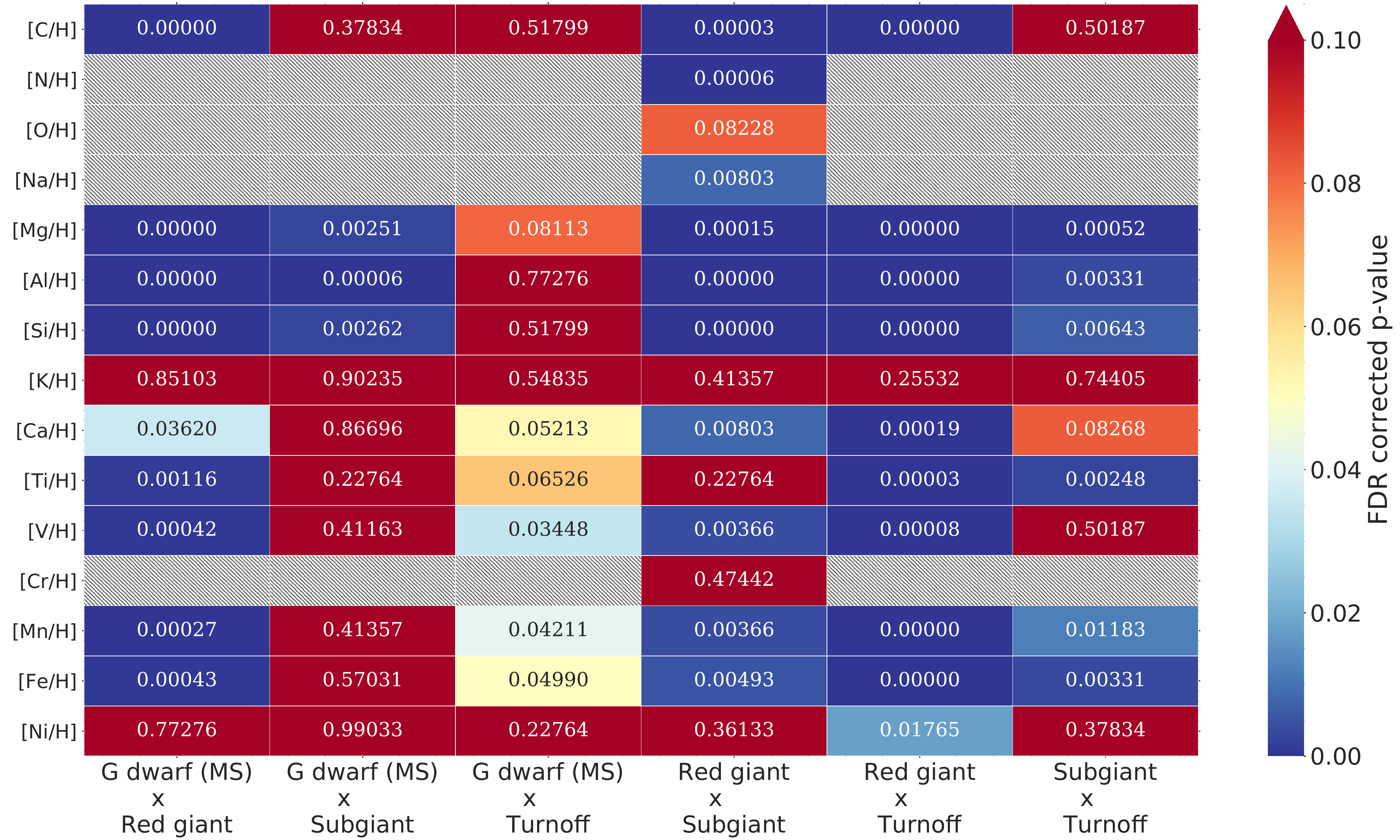}
\caption{The Kolmogorov--Smirnov test for the different elements ($y$-axis) on the different stellar classes ($x$-axis). The obtained $p$-value is color coded from blue to red. The red colors are saturated at 0.1.}
\end{center}
\label{fig_KStest_a}
\end{figure*}

In Figure \ref{fig_KStest_a} we present the results of the Kolmogorov--Smirnov two-sided test (K-S test) comparing the individual abundances for each stellar class.
The vertical axis represents the [X/H] derived here, and the horizontal axis represents the subgroups being compared. 
Each cell shows the $p$-value of the K-S test and is colored as shown in the side color bar. We designed the color scale to give a blue color if the samples are clearly distinct, a yellow color if the $p$-value is near to 0.05, and a red color if we cannot reject the null hypothesis, i.e., the samples are not distinguishable.
Note that we have applied False Discovery Rate (FDR, \citealt{Benjamini1995}) correction in order to account for the fact that we are performing many hypothesis tests simultaneously and spurious rejections of the null hypothesis are therefore expected.  
Regardless of the threshold that we use, we obtain outstanding segregation for red giant and turnoff stars based on their abundances. The K abundance is the one with higher $p$-values ($>$ 0.03) for all scenarios.
On the other hand, the two classes most difficult to separate based on their abundances are the main-sequence and the turnoff stars. 
The abundances of Mg, Ca, V, and Fe are the best ones to distinguish between these classes.
The Mg abundances show significant differences among all stellar classes (with $p$-values $<$ 0.10 for all comparisons).

\subsubsection{As a Function of Stellar Parameters}

\begin{figure*}
\begin{center}
\includegraphics[width=0.9\linewidth]{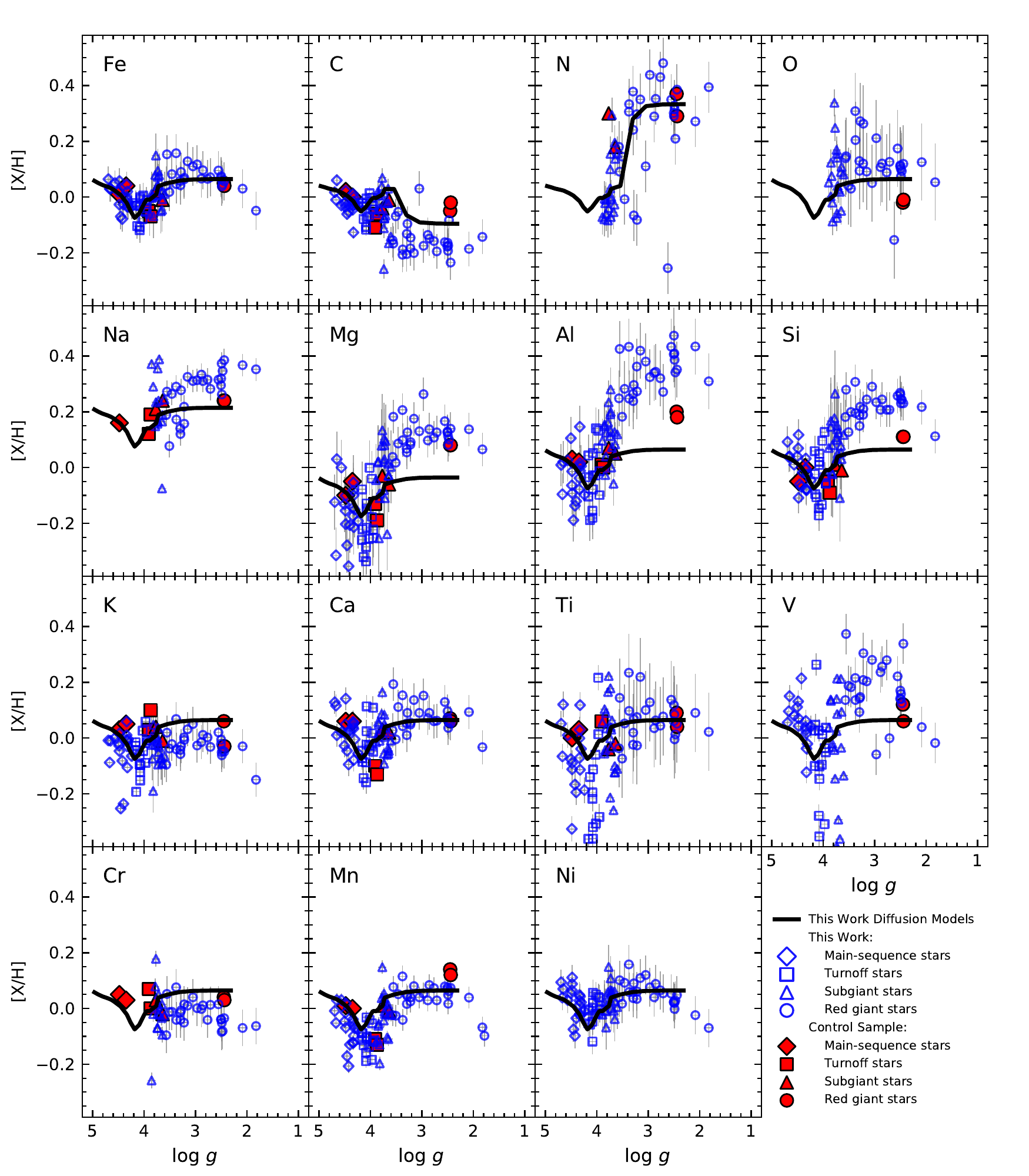}
\caption{The chemical abundances for the studied stars are shown as a function of log $g$. The symbol notation is similar to Figure \ref{fig1_diagram} (open symbols instead of filled symbols).
The diffusion models calculated in this work are shown as solid black lines.}
\end{center}
\label{Axlogg}
\end{figure*}

In Figure \ref{Axlogg} we display the derived individual abundances as a function of surface gravity for the fifteen elements studied.
We use the same symbol notation as in Figure \ref{fig1_diagram}, but with open symbols instead of filled.
We also show the line-by-line manual abundance results from \cite{Souto2018}, our control sample.
Atomic diffusion models computed for this work (see Section 6) are over-plotted for each element (C and N including mixing processes). 
We note that the diffusion models for Na and Mg abundances were slightly shifted in order to better fit the observed abundances. 

From visual inspection ---and in agreement with the results from the K-S test--- we can organize the element variations as a function of surface gravity (as well as $T_{\rm eff}$ and $M_{\star}$) into three groups of elements: \textit{(i)} C and N, with abundances displaying a different behavior for the evolved subgiant and red giant stars (as a consequence of dredge-up mechanisms); \textit{(ii)} O, Na, and Cr as their abundances are not reliable for the main-sequence and turnoff stars since their spectral lines become too weak; \textit{(iii)} the elements showing a dip, either sinuous or small, in the elemental abundance close to log $g$ = 4.00 dex (Fe, Mg, Al, Si, K, Ca, Ti, V, Mn, and Ni)
The derived abundances of Mg, Al, and Si, present the most significant changes between the stellar classes (excluding N), where the red giant abundances are 0.10 to 0.20 dex higher than those from the subgiants.

\begin{figure*}
\begin{center}
\includegraphics[width=0.9\linewidth]{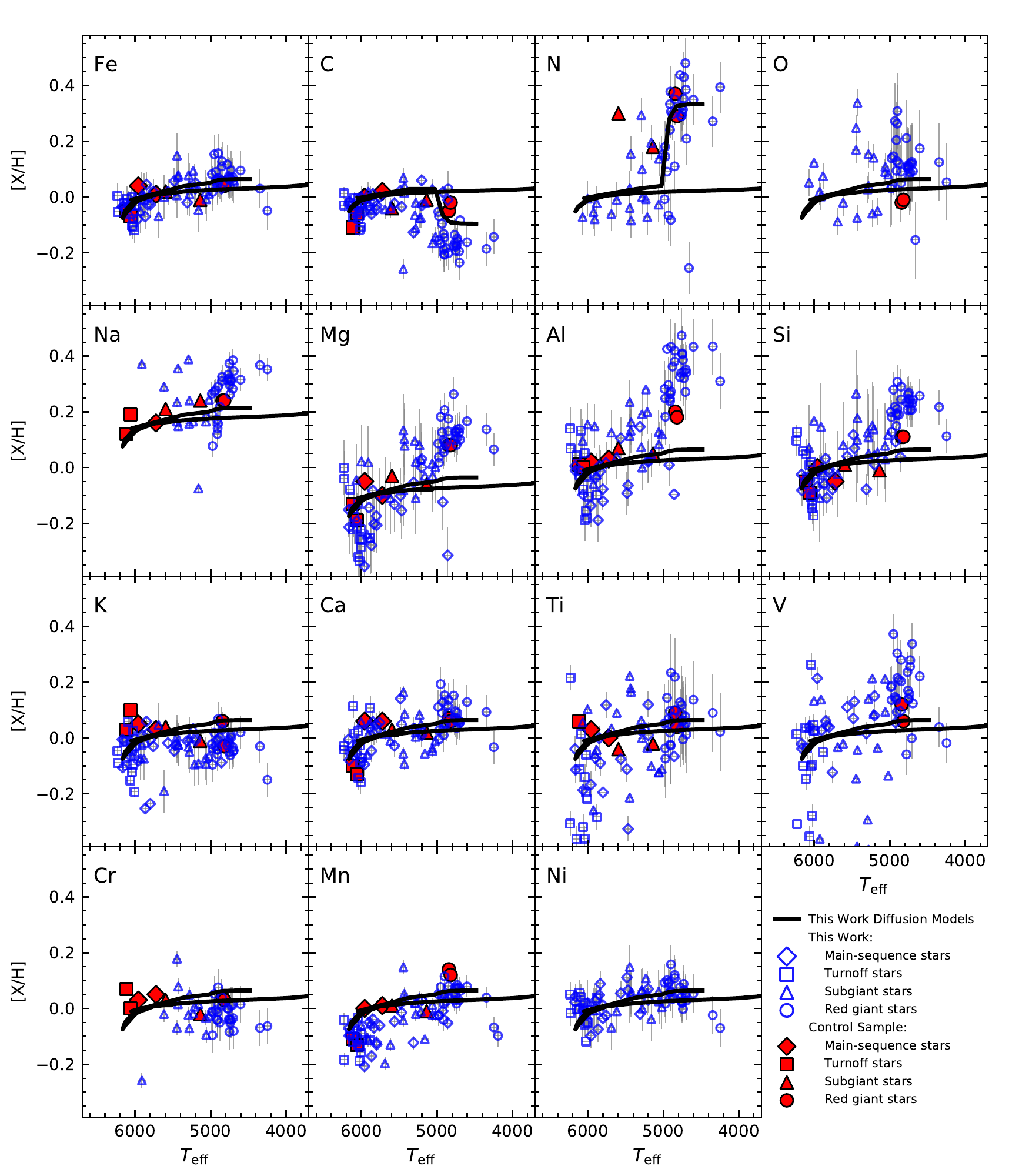}
\caption{Same as Figure \ref{Axlogg}, except shown as function of $T_{\rm eff}$.}
\end{center}
\label{AxTeff}
\end{figure*}

In Figure \ref{AxTeff}, we present the abundance results as a function of $T_{\rm eff}$ in M67 stars, with diffusion models also shown.
Overall, the behavior seen in Figure \ref{AxTeff} indicates an abundance increase (in the range 0.00--0.40 dex) as $T_{\rm eff}$ decreases from 6000 K to 4000 K.
The elements showing a smooth increase or decrease in abundance as functions of $T_{\rm eff}$ are Fe, Ca and, Mn.
The elements most sensitive to $T_{\rm eff}$: Na, Mg, Al, and Si, show a monotonic increase in their individual abundances. 
Similar to the trends with log $g$, C shows a particular behavior and the abundance variation of N shows a maximum value around $T_{\rm eff}$ $\sim$ 4700 K and then decreases for higher and lowers values of $T_{\rm eff}$. 
The elements presenting the least sensitivity to $T_{\rm eff}$ are K, Cr and, Ni.
Ti and V show the most significant abundance scatter in the analysis as a function of both log $g$ and $T_{\rm eff}$.

\cite{Souto2018} showed that atomic diffusion processes can explain the abundance variations of M67 stars across the different evolutionary stages.
However, other physical processes are also relevant in the context of abundance variations, where the most significant sources of deviations, not precisely in order, are: non-LTE effects, 1-D or 3-D treatment of the model atmosphere, stellar rotation (v sin $i$), mixing process (e.g., first dredge-up), and atomic diffusion processes. In the following sections, we discuss the impact of these possible deviations in our results.

\section{Possible Explanations to the Abundance Trends}

Figures \ref{Axlogg}, and \ref{AxTeff}, show significant abundance variations as a function of the stellar parameters (log $g$ and $T_{\rm eff}$). Such abundance trends are not expected to occur in open clusters ---due to the homogeneity of the stars formed by the same material--- unless some additional effect/mechanism is playing a role in the stellar atmosphere, or in the abundance determination itself.

\subsection{Non-LTE Deviations in the NIR}

Deviations from the local thermodynamical equilibrium have been studied mostly at optical wavelengths where strong deviations are found to occur in metal-poor evolved red giant stars (\citealt{Asplund2005nonLTE,Asplund2009}). 
In the NIR, in particular in the H-band, the works of \cite{Cunha2015} and \cite{Zhang2016,Zhang2017} have investigated non-LTE effects in Na I, Mg I, and Si I lines in the APOGEE spectra, finding deviations from non-LTE in these elements to be usually smaller than 0.05 dex (see also the discussion in \citealt{Souto2018}).
Using the results from \cite{Bergemann2008Mn} and \cite{Bergemann2012FeTi,Bergemann2013Si,Bergemann2015Mg} compiled from a Maria Bergemann web site (nlte.mpia.de), we created a grid of non-LTE deviations for five elements: Fe, Mg, Si, Ti, and Mn.
The deviations were estimated for each stellar class, assuming a solar metallicity and $T_{\rm eff}$ = 4700 K, log $g$ = 2.40, and $\xi$ = 1.60 km s$^{-1}$ for red giants, $T_{\rm eff}$ = 5400 K, log $g$ = 3.70, and $\xi$ = 1.25 km s$^{-1}$ for subgiants, $T_{\rm eff}$ = 6100 K, log $g$ = 3.90, and $\xi$ = 1.15 km s$^{-1}$ for turnoff stars, and $T_{\rm eff}$ = 5850 K, log $g$ = 4.40, and $\xi$ = 1.00 km s$^{-1}$ for main-sequence stars.
We adopted 1-D plane-parallel models computed with MAFAGS-OS for all stellar classes. 
In Table 5 we summarize the average non-LTE correction for each stellar class and element.

\begin{figure*}
\begin{center}
\includegraphics[width=1\linewidth]{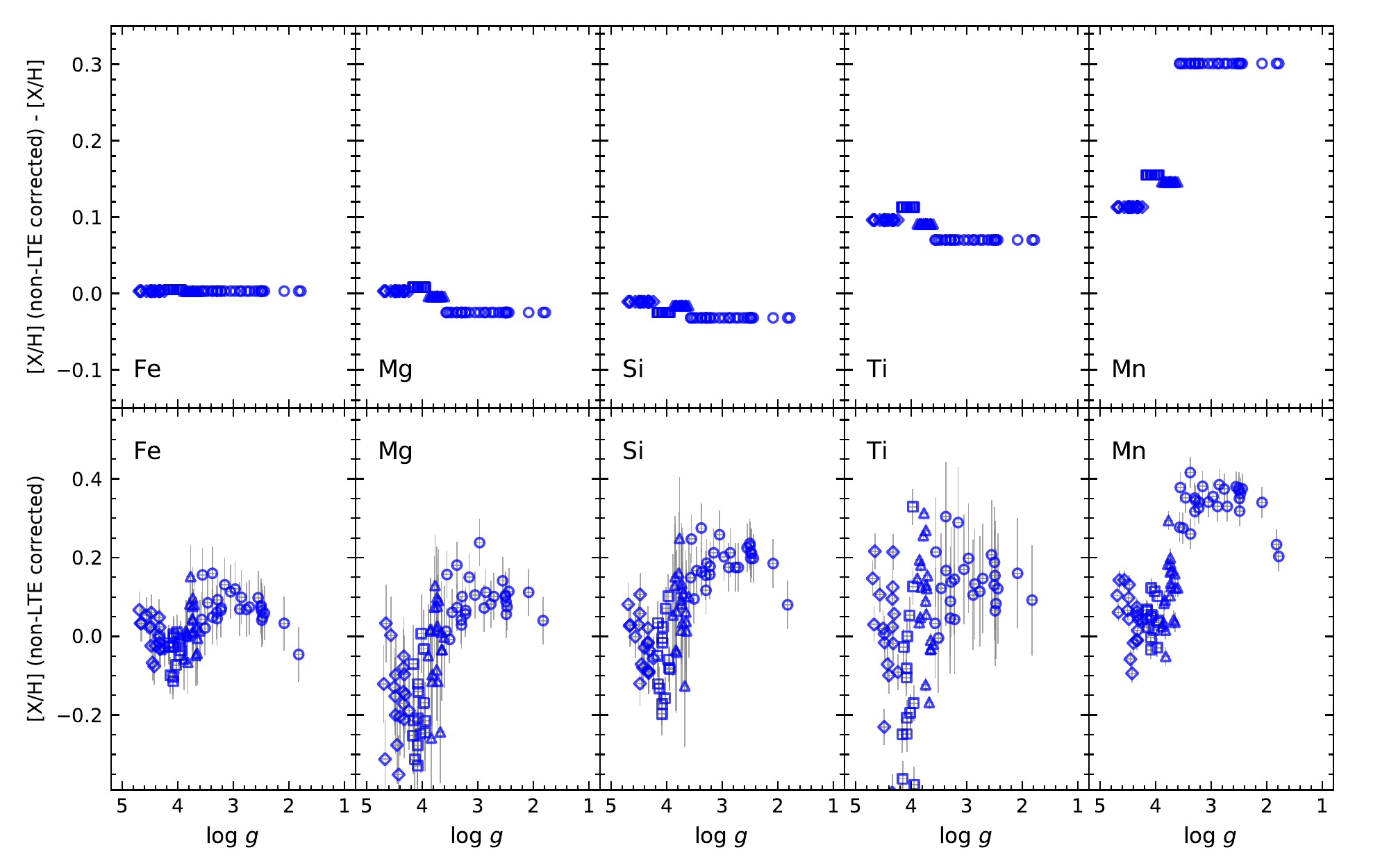}
\caption{Top panel: log $g$ \textit{versus} [X/H] (corrected-derived) individual abundances from non-LTE deviation. Bottom panel: log $g$ versus [X/H] corrected from non-LTE deviations. In both panels, the symbols follow the same notation as Figure \ref{Axlogg}.}
\end{center}
\label{fig_nonLTE}
\end{figure*}

In Figure \ref{fig_nonLTE} we show the non-LTE corrected abundances for five elements studied (Fe, Mg, Si, Ti, and Mn).
The top panel displays the abundance differences from [X/H]$_{non-LTE}$ - [X/H]$_{LTE}$, and in the bottom panel we show a similar plot as Figure \ref{Axlogg}, but now using the [X/H]$_{non-LTE}$.

The iron abundances do not show significant non-LTE deviations, as seen in Table 5, where $\delta$(nonLTE-LTE) are smaller than 0.01 dex for all stellar classes.
For Mg and Si, the deviation is very similar for main-sequence stars, both positive, being almost null for Mg. 
For subgiant and red giants stars, we obtain small negative non-LTE corrections. 
The deviations for Ti and Mn are more significant in this study. 
For Ti, the deviations are positive for the stellar classes studied here, with the major deviation observed in turnoff stars ($\delta$(nonLTE-LTE) = 0.11 dex).
When applying non-LTE corrections, we do not see a strong change in the abundance \textit{vs} log $g$ diagram, when compared to the LTE one presented in Figure \ref{Axlogg}. 
The abundances of Ti are shifted in all classes, resulting in a higher scatter as a function of log $g$.
The Mn corrections show the most significant differences, $\delta$(nonLTE - LTE) $\sim$ 0.13 dex for main-sequence, turnoff, and subgiants, and $\delta$(nonLTE - LTE) $\sim$ 0.30 dex for red giants.
The inclusion of non-LTE corrections in the analysis does not erase the observed abundance trends in the different stellar classes.

\begin{deluxetable}{lccccc}
\tablenum{5}
\tabletypesize{\tiny}
\tablecaption{Non-LTE corrections}
\tablewidth{0pt}
\tablehead{
\colhead{Stellar Class} &
\colhead{Mg} &
\colhead{Si} &
\colhead{Ti} &
\colhead{Mn} &
\colhead{Fe}}
\startdata
Main-sequence 	&	+0.003	&	-0.011	&	+0.096	&	+0.113	&	+0.003\\
Turnoff 	&	+0.008	&	-0.025	&	+0.113	&	+0.155	&	+0.005\\
Subgiant 	&	-0.004	&	-0.016	&	+0.091	&	+0.146	&	+0.003\\
Red giant 	&	-0.025	&	-0.032	&	+0.070	&	+0.301	&	+0.003\\
\tablewidth{0pt}	
\enddata
\end{deluxetable}

\subsection{1-D or 3-D Model Atmospheres}

Stellar atmospheres are 3-D and time-dependent; however, by convenience, we usually treat model atmospheres as having 1D plane-parallel or spherical geometry in hydrostatic equilibrium.
This approximation simplifies the analysis, but can lead to systematic errors in the derived quantities (atmospheric parameters or chemical abundances).

The use of a 1-D treatment of the stellar atmosphere requires the inclusion of ``ad hoc'' parameters to account for velocities that broaden the profiles at microscopic (microturbulence) and macroscopic (macroturbulence) levels.
A precise determination of the microturbulence parameter minimizes the deviations from the results obtained with 3D models.

As in non-LTE studies, 3-D effects are also transition dependent, and analyses for NIR H-band transitions have been limited. 
The studies of \cite{Asplund2005nonLTE,Asplund2009}, \cite{Caffau2011} have summarized various effects and corrections for elemental abundances using optical spectra as a reference. 
In this Section, we will summarize these effects for solar metallicity stars to verify whether the abundance trends discussed in Section 4 could be explained by 3-D effects.

\cite{Caffau2011} determined solar abundances from a 3-D non-LTE analysis using the CO$^{5}$BOLD code, providing 3D abundance corrections for several elements. 
For Fe, \cite{Caffau2011} find 3-D corrections to be about 0.03 dex using the solar spectrum.
The C abundance reported by \cite{Caffau2011} has a -0.02 dex 3-D correction, while for K the authors obtain a correction of 0.07 dex.
From \cite{Caffau2009} the solar N abundance is reported to have a 3-D correction smaller than 0.01 dex.

The previous work by \cite{Bergemann2012} studied the 3-D deviations for stars in different evolutionary stages at different metallicities. They find 3-D effects in the iron abundance for the Sun to be very small: 3-D corrections $\sim$ 0.01 dex.
More recently, \cite{Bergemann2017} studied the Mg abundances in the Sun and found 3-D corrections to be $\sim$0.02 dex. \cite{Amarsi2017} studied non-LTE 3-D Si abundances in the Sun and found corrections to be lower than 0.01 dex.
\cite{Amarsi2016} analyzing the O I forbidden line at 630nm find 3-D corrections to the O abundance to be between 0.05--0.20 dex, negative in the Sun and reaching higher values for turnoff stars.

All 3-D corrections discussed above are smaller than 0.05 dex (except for K), which is at the limit of the measurement uncertainties of this work.
Given the small 3-D corrections found for main-sequence stars, as well as, the lack of studies in the literature for turnoff, subgiants and red giants stars at solar metallicity, we conclude that deviations from 3-D modeling are not enough to explain the abundance trends observed in this work.

\subsection{Stellar Rotation}

The study of the relation between stellar rotation and abundance variations in late-type stars is often motivated by the investigation of lithium depletion. Several authors have found correlations between  stellar rotation and the lithium abundance depletion, such as \cite{Balachandran1990}, \cite{Balachandran1995}, \cite{King2000}, \cite{dasilva2009}, \cite{CantoMartins2011}, and \cite{DelgadoMena2014}.
None of the spectra analyzed in this study exhibit measurable rotational broadening (vsin($i$)) above the limits set by the APOGEE spectral resolution of $\sim$ 7--8 km s$^{-1}$.

\section{Discussion}

The abundance results obtained for M67 stars show evidence that both mixing and atomic diffusion are operating, thus stellar evolution models that include diffusion will be compared to the observationally derived abundances .

\subsection{Stellar Evolution Models}

We computed our mixing and atomic diffusion models using solar models (solar metallicity and solar age 4 Gyr) to calibrate the degree of gravitational settling precisely (using the surface solar He as a proxy); this gives a predicted reduction in the efficiency of the settling of 15\%, or an effective coefficient of 0.85. 
The methodology adopted in the modeling of mixing and atomic diffusion is described in detail in \cite{Bahcall2001} and \cite{Delahaye2006}.
We note that, overall, our models agree with the ones from MIST (\citealt{Choi2016}, \citealt{Dotter2017}); however, our models cover all the species studied in this work, while the MIST models are not available for Al, K, V, Cr, Mn, and Ni.

\subsection{Mixing Processes: First Dredge-up (FDU)}

When a low-mass star, such as a $\sim$1.2M$_{\odot}$ M67 star that is currently evolving off of the main-sequence and across the subgiant branch, reaches the base of the red giant branch (RGB), the outer convective envelope reaches its largest extent in mass.  At this point in the H-R diagram (where T$_{\rm eff}\sim$5000K and log $g$ $\sim$3.5 in M67), the base of the convective envelope ingests material that has been exposed previously to H-burning via the CN cycle.  As a consequence of CN-cycle H-burning, this nuclear-processed material contains an enhanced abundance of $^{14}$N and a decreased abundance of $^{12}$C. The convective envelope will carry this mixture to the surface, resulting in a lower surface abundance of $^{12}$C and a larger abundance of $^{14}$N for stars evolving onto the RGB; this phase of stellar evolution is referred to as first dredge-up, or FDU (\citealt{Iben1965}; for a more recent overview of the various red giant dredge-up episodes see \citealt{KarakasLattanzio2014}).
In the case of dredge-up in M67 red giants, the $^{14}$N abundance is predicted to be enhanced by roughly $\sim$+0.30 to +0.40 dex, while the $^{12}$C abundance is predicted to be depleted by $\sim$-0.10 to -0.20 dex.  The magnitudes of the abundance changes in C and N are a function of red giant mass (\citealt{Iben1965}), with larger mass stars having deeper convective envelopes which dredge up more nuclear-processed material, resulting in larger $^{14}$N enhancements and larger $^{12}$C depletions, producing lower C/N ratios. 

The expected relationship between red giant mass and C/N ratio has been exploited by a number of recent studies using APOGEE data and results, e, g., \cite{Martig2016}, \cite{Ness2016} (see also Feuillet et al. \textit{in preparation}) to produce age--mass relatios as a function of red giant [C/N] abundances, while 
\cite{Masseron2015} have analyzed [C/N] to study the possible formation of the thin and thick disk.

In addition to standard convection in 1D, other physical processes can modify the interior abundance profiles in stars as they evolve from the main-sequence, across the subgiant branch, and onto the red giant branch, with two important processes being rotation and the inversion of the mean molecular weight gradient in a small region outside of the H-burning shell created by $^{3}$He-burning via $^{3}$He($^{3}$He,2p)$\alpha$ (\citealt{Eggleton2006}; \citealt{CharbonnelZahn2007}): this last process is referred to as thermohaline mixing. 
The inclusion of rotation-induced mixing and thermohaline mixing produces larger carbon depletions and larger nitrogen enhancements as a result of FDU. In this Section, we use $^{12}$C and $^{14}$N abundances derived here to compare with predictions from various models of first dredge-up mixing.

\begin{figure*}
\begin{center}
\includegraphics[width=0.44\linewidth]{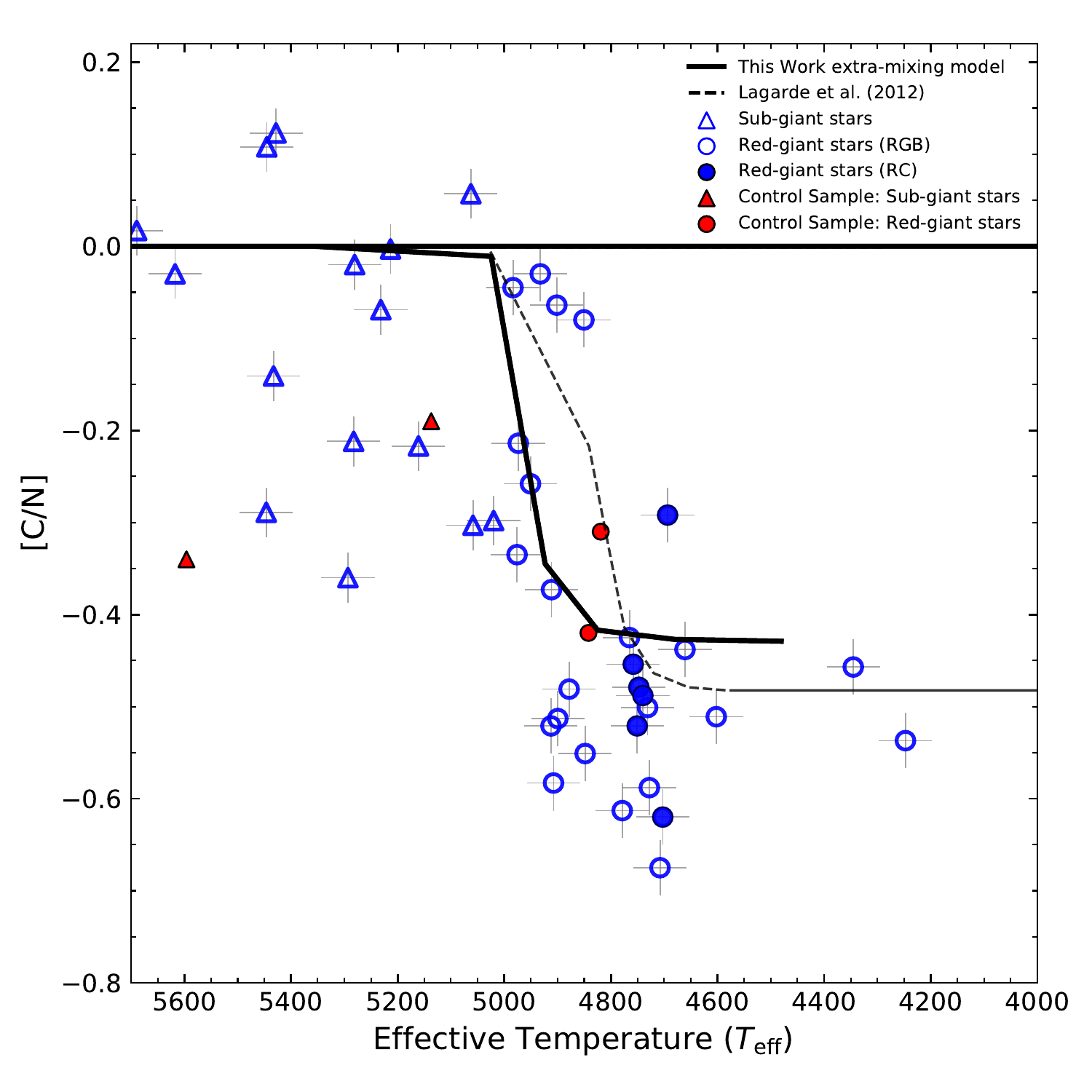}(a)
\includegraphics[width=0.44\linewidth]{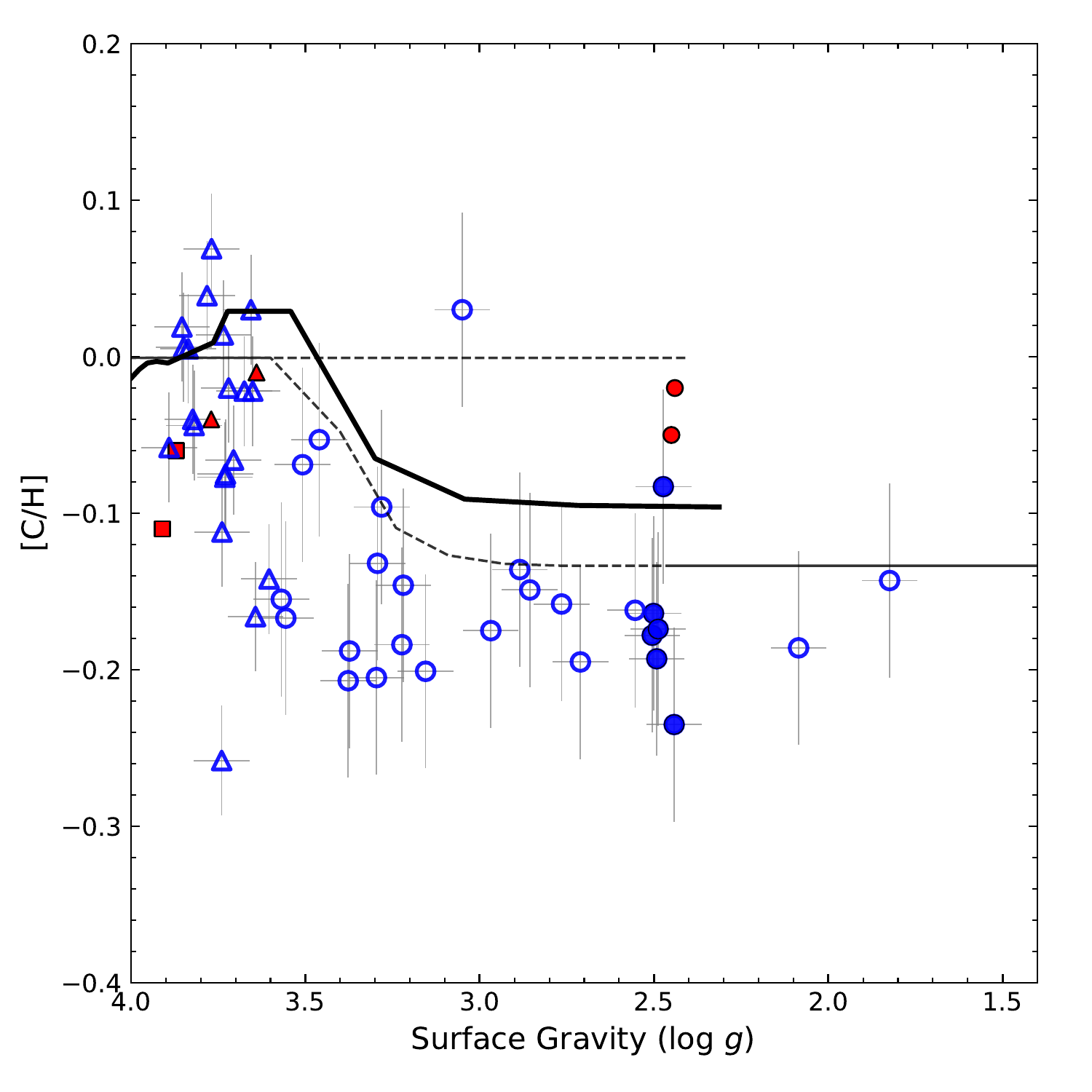}(c)
\includegraphics[width=0.44\linewidth]{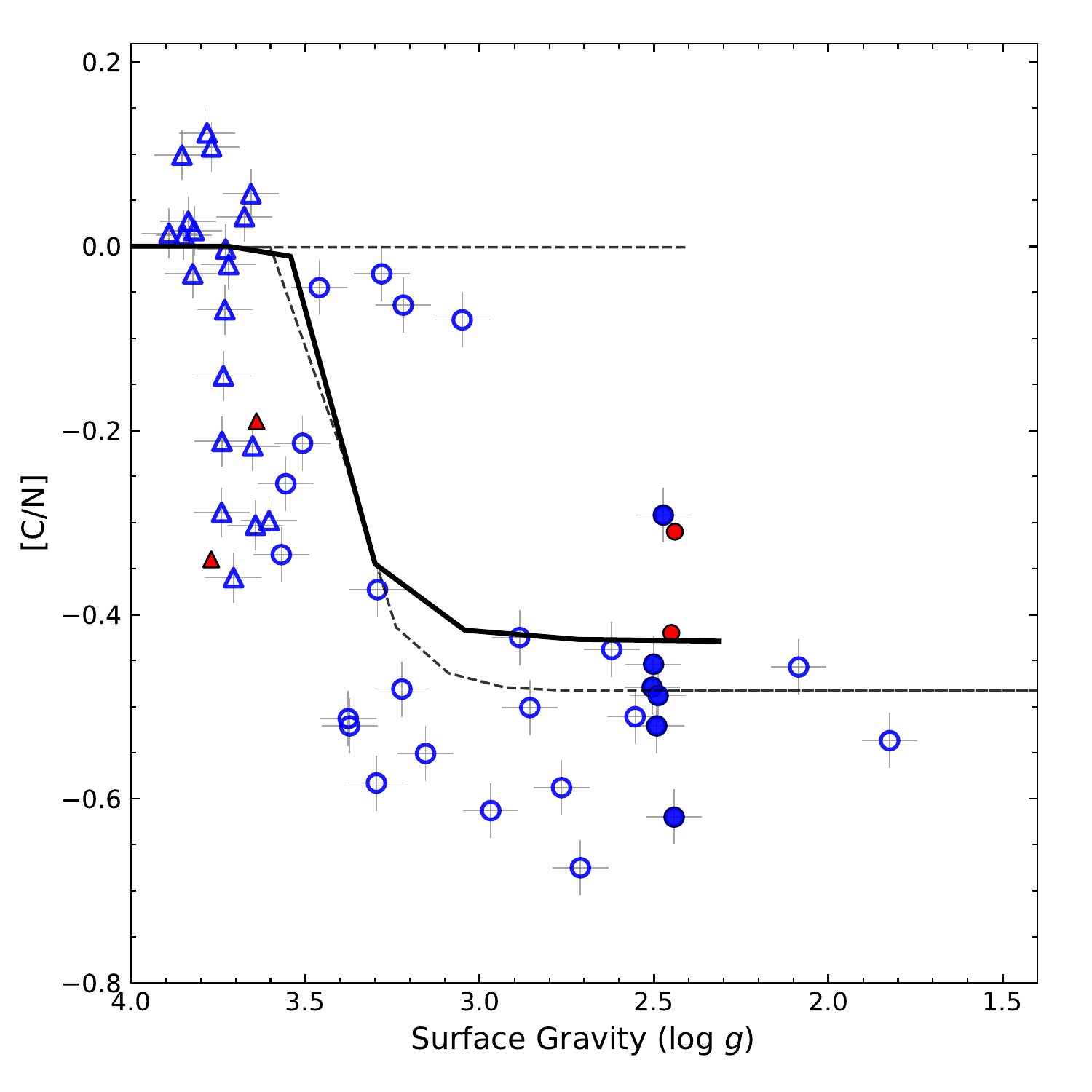}(b)
\includegraphics[width=0.44\linewidth]{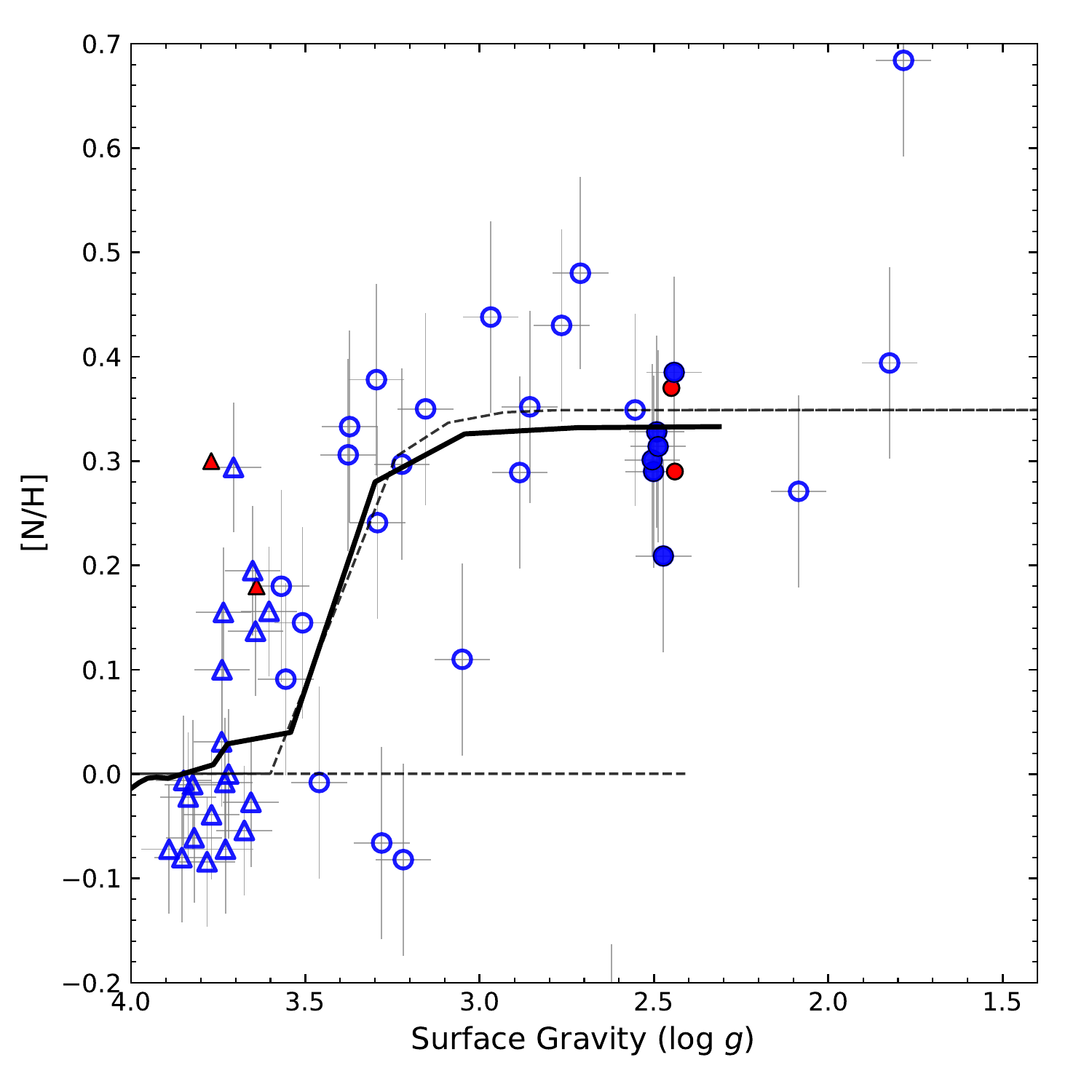}(d)
\caption{
Panels a and b shows the $T_{\rm eff}$ $x$ [C/N] and the log $g$ $x$ [C/N], respectively. Panels c and d present the [C/H] $x$ log $g$ and [N/H] $x$ log $g$ diagrams. We display the mixing models from this work (solid black curve) and from \cite{Lagarde2012} (dashed black curve) as a comparison. 
All symbols follow the same notation as Figure \ref{Axlogg}, with the inclusion of filled circles for the red clump stars.}
\end{center}
\label{fig_dredgup}
\end{figure*}

As shown in Figure \ref{fig_dredgup}, the M67 red giants display clear evidence of the first dredge-up through the behavior of the C and N abundances as functions of both T$_{\rm eff}$ and log $g$ (which map the position of a star along the subgiant and red giant branches); observed APOGEE abundances are plotted as the various symbols, while models are plotted as the continuous lines and are models from this study, along with those from \cite{Lagarde2012}.  
The left panels of Figure \ref{fig_dredgup} plot the [C/N] values versus T$_{\rm eff}$ (top, a) and log $g$ (bottom, b), with the [C/N] values decreasing rapidly at T$_{\rm eff}\sim$5000K and log $g$$\sim$3.5, right at the base of the RGB as predicted by FDU. The right panels plot the individual abundances of $^{12}$C (as [C/H]) and $^{14}$N (as [N/H]) versus log $g$. 
Carbon and nitrogen abundance differences between red giants on the RGB relative to those in the RC were found to agree with results from \cite{Tautvaisiene2000} and \cite{Masseron2017}, who found slightly lower values of C/N in RC stars compared to those on the RGB. Our values for M67 stars are $\langle$$^{12}$C/$^{14}$N$\rangle$$_{\rm RGB}$ = 1.86 and $\langle$$^{12}$C/$^{14}$N$\rangle$$_{\rm RC}$ = 1.40, excluding the two evolved stars with log $g$ $<$ 2.1 dex, which places them on the upper RGB or possibly in an early-AGB phase of evolution.

Figure \ref{fig_dredgup} also highlights differences in the C and N abundance variations predicted from mixing models when compared to those abundances derived in this study.
In the left panels of Figure \ref{fig_dredgup} (a and b), we show the [C/N] ratio as a function of $T_{\rm eff}$ and log $g$, respectively, and note that the overall observational results follow the model predictions, although the observed [C/N] values are systematic lower.
Such a difference can be a consequence of an overestimated nitrogen abundance in our analysis (as pointed out by \citealt{BertelliMotta2017} using ASPCAP data), due to a sub-estimated log $g$. 
In the right panel of Figure \ref{fig_dredgup}, we present the [C/H] (panel c) and [N/H] (panel d) abundances as a function of log $g$. 
For nitrogen, the abundances are in agreement with the models; however, the observational carbon abundances differ from the models by $\sim$-0.15 dex. 
We conclude that the abundance variations observed for $^{12}$C and $^{14}$N in the subgiant and red giant stars can be explained well by FDU mixing models.
The mixing models here (as well as from \citealt{Lagarde2012}) predict changes for the other elemental abundances to be smaller than 0.01 dex as the star evolves. Therefore, mixing models cannot explain their abundance variations.

\subsection{Atomic Diffusion}

Atomic diffusion is a likely explanation for most of the observed abundance variations across the H-R diagram in M67, thus adding members of this old open cluster to those stars in which diffusion has been observed. 
Evidence of diffusion in the Sun is found both in its surface helium abundance, which is lower than the initial value, as well as the solar sound speed profile being best fit by models that include diffusion (\citealt{Bahcall1995};\cite{Chaboyer1995}).
Lithium abundances settle at a rate similar to He, and the flatness of the Spite Li plateau is likely set by diffusion (\citealt{Chaboyer1992}).  The diffusion signature can be altered or erased by mixing, for example, mixing driven by rotation and dredge-up (see Sections 5.3 and 5.4), thus complicating the detection and interpretation of diffusion patterns.  Such mixing processes are likely at work in the Sun, which has a smoother composition profile than that predicted by diffusion alone, with the magnitude of diffusion being overestimated by about 25\%.  This is also confirmed by looking at A-type stars -- if they rotate fast enough, they are not chemically peculiar (\citealt{Michaud1970}, see also \citealt{Michaud2015}).
The interplay between diffusion creating abundance signatures that various mixing processes can then modify means that there are not necessarily firm theoretical predictions about the amplitude of the diffusion signature and its mass or metallicity dependence. Reasonably well-motivated trends can be expected, though, and \cite{Chaboyer1995b, Chaboyer1995c}, \cite{Choi2016}, \cite{Dotter2016}, \cite{Dotter2017} have approximated limits on how efficient diffusion can be in thin surface convective zones.

A few previous studies have probed atomic diffusion in cluster stars, with most of them focused on low-metallicity globular clusters: \cite{Korn2007}, \cite{Lind2008}, and \cite{Nordlander2012}. The latter analyzed stars belonging to the globular cluster NGC 6397, with a metallicity of [Fe/H] = -2.00 and age of 13.5 Gyr, with their sample containing stars from the turnoff point (TOP) up to the red giant branch (RGB).
The abundances of Li, Mg, Ca, Ti, Cr, and Fe in those studies were derived from high-resolution optical spectroscopy, and they found that changes in the stellar abundances for different evolutionary phases are in good agreement with predictions from diffusion models from the literature, see \cite{Richard2002,Richard2005}. 
In particular, \cite{Nordlander2012} found abundance differences of 0.06 and 0.18 dex between TOP and RGB stars in NGC 6397, with the largest difference for Mg, and which is a much smaller variation than we see in M67, for example for Mg or Al. 
Of course the chemical abundance of NGC 6397 is rather distinct from that of M67.

\cite{Onehag2014} found some evidence of atomic diffusion operating in M67. 
This was later supported by the abundance results in \cite{Blanco-Cuaresma2015}.
As previously mentioned, abundance differences of up to $\sim$0.20 dex between the turnoff and red giant stars were observed by \cite{Souto2018} for the elements O, Na, Mg, Al, Si, Ca, V, Mn, and Fe.
A comparison of the results in \cite{Souto2018} with the diffusion patterns from the models by \cite{Choi2016} and \cite{Dotter2017} indicated good agreement.
In addition,
\cite{BertelliMotta2018} found $>$0.15 dex abundance differences from main-sequence to red giant stars for the elements Al, Si, Mn, and Ni. \cite{Gao2018} found a good match between the abundance variations for Al and Si with diffusion models.

\subsection{Atomic Diffusion in M67 Stars}

We find significant abundance differences (up to $\sim$ 0.50 dex) for most of the studied species between main-sequence, turnoff, subgiant, and red giant stars in M67.
Using the K-S test (Figure \ref{fig_KStest_a}), we obtained clear evidence of abundance differences between stars in different evolutionary stages in M67. In addition, we showed (Section 4) that the abundances of stars belonging to the same evolutionary class are indistinguishable. 

\begin{figure*}
\begin{center}
\includegraphics[width=0.99\linewidth]{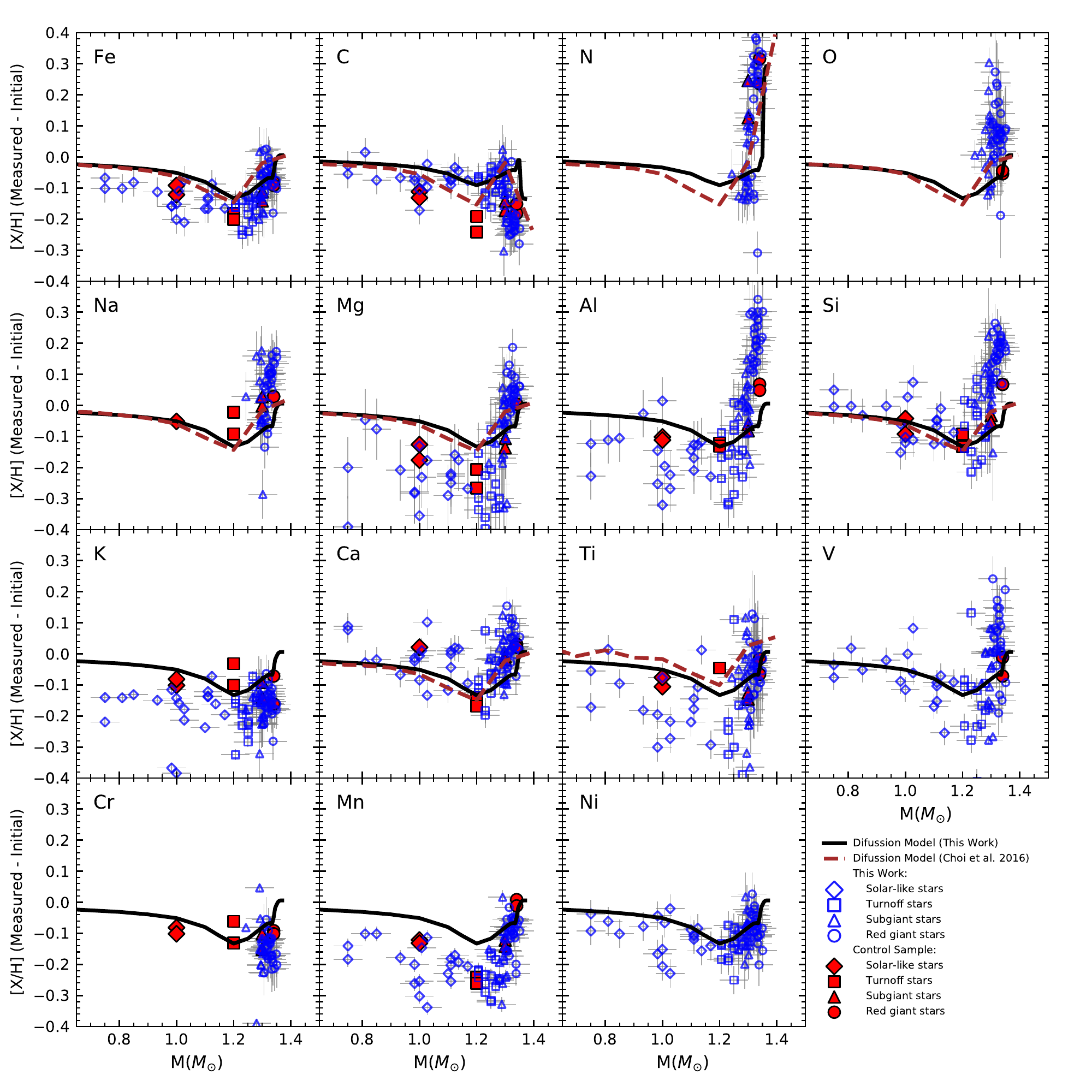}
\caption{Diagram of the stellar mass as a function of $\Delta$[X/H], where  $\Delta$[X/H] indicate the derived metallicity from the stellar photosphere minus the initial cluster composition. 
The black and brown lines shows the atomic diffusion models from this work and MIST, respectively.
All symbols follow the same notation as Figure \ref{Axlogg}.
}
\end{center}
\label{fig_deltabuxmass}
\end{figure*}

In Figure \ref{fig_deltabuxmass} we present the mass--$\Delta$[X/H] ([X/H]$_{\rm Current}$ - [X/H]$_{\rm Initial}$) diagram for the twelve studied elements. Similar diagrams with abundances as a function of surface gravity and effective temperature are presented in Figures 5 and 6. In all three figures, we show the atomic diffusion models computed in this work as solid black lines and the MIST models as brown dashed lines.
The pristine Fe abundance in the models is assumed to be the mean Fe abundance for the red giants, which is used as the fiducial point (i.e., $\delta$[Fe/H] = 0.00) for the initial cluster value.
We note that all other abundance ratios are assumed solar, i.e. [X/Fe] = 0.00.
The abundance variations across the H-R diagram indicate that atomic diffusion is operating in most of the studied elements. 
The models for all the elements display similar trends driven by atomic diffusion, except for C and N, which include mixing signatures. 

The complex trend observed in the carbon abundances is a consequence of diffusion operating in the main-sequence and turnoff stars (smaller convective envelopes), and mixing at the first dredge-up being responsible for the carbon abundance variation in subgiant and red giant stars (Section 6.2; Figure \ref{fig_dredgup}).
These results suggest that atomic diffusion dominates over mixing in the main-sequence and turnoff stars, while mixing processes control the abundance changes in subgiant and red giant stars. 

The nitrogen abundance variation can be explained as a signature of first dredge-up (Section 6.2). 
For oxygen, the scattered abundance results for red giant and subgiant stars, combined with the lack of results for main-sequence and turnoff stars, impedes detecting signatures of diffusion. 
Due to the weakness of CN and OH molecular lines in the APOGEE spectra of main-sequence and turnoff stars, it is not possible to derive N and O abundances in such stars.

The comparison of the abundance patterns for all elements with the model predictions indicates an overall good match between the atomic diffusion models and the derived abundances across the H-R diagram. 
However, the derived abundances exhibit a more significant dip across the main-sequence---turnoff when compared to what is expected from the atomic diffusion models, in particular for Mg, Al, Ti, and Mn. 

For Al, Mg, Si, and to a lesser degree V, the relative dip across the main-sequence---turnoff stars is more significant because the red giant abundances are higher than those predicted by the models. (The Na abundances of red giants are also higher than the models, but there are no abundances for turnoff and main-sequence stars.)
On the other hand, for Ti and Mn the dip is more considerable because the abundances of turnoff and main-sequence stars are lower. 
As discussed in Section 5 (Figure \ref{fig_nonLTE}), non-LTE corrections for Mg, Si (as well as Fe) would reduce the abundance dip by a factor of $\sim$ 0.03 dex, while for Ti, the dip would be reduced by roughly 0.05 dex. 
The non-LTE corrections for Mn, on the other hand, would systematically change the red giant abundances and increase the abundance difference between turnoff and red giant stars by $\sim$ 0.14 dex, which would worsen the comparison with the models.

\section{Summary}

Given its combination of age and metallicity in addition to the numerous detailed studies in the literature, M67 remains a prime cluster to test for not well-understood physical/chemical processes in stellar spectroscopy.

In this paper, we present individual abundances of fifteen elements (C, N, O, Na, Mg, Al, Si, K, Ca, Ti, V, Cr, Mn, Fe, and Ni) derived from a 1-D LTE analysis of 83 stars in M67.
The abundances were obtained via chi-square minimization of the high-resolution SDSS-IV/APOGEE spectra with the qASPCAP code.
The stellar sample is composed of stars in different evolutionary stages (19 main-sequence; 15 turnoff; 20 subgiant; and 29 red giants) with the aim to quantify abundance trends across the different stellar evolutionary phases (\citealt{Souto2018}, \citealt{Onehag2014}, \citealt{BertelliMotta2018}, and \citealt{Gao2018}).

We obtain significant abundance differences (of up to 0.30-0.40 dex) as a function of stellar parameters ($T_{\rm eff}$, log $g$, and mass), which map the different stellar evolutionary classes. 
Studying the abundance variations within the same stellar classes, we find a much lower scatter, of about $\sim$ 0.05 -- 0.10 dex.
Using the K-S test, it is found that the abundances within each stellar class are indistinguishable; while performing the test for the different classes, we obtain clear segregations for the red giant and turnoff stars in most of the elements analyzed. 

We compiled non-LTE corrections for Fe, Mg, Al, Si, and Mn, finding them to be small for all elements ($<$ 0.10 dex), except for Mn, which were between 0.15--0.30 dex in all stellar classes analyzed. 
With the use of non-LTE corrections to our derived abundances, we still observe clear abundance trends across the H-R diagram.

We found that mixing models explain well the abundance variations of C and N for subgiants and red giant stars. 
We see atomic diffusion operating in the C abundances in stars from the main-sequence--turnoff point. 
The atomic diffusion models computed in this work (as well as from the literature) predict reasonably well the remaining abundance patterns for the stars at different evolutionary stages studied in this work, and therefore, we conclude that atomic diffusion operates in M67, more efficiently in the turnoff stars and in most of the elements (C, Mg, Al, Si, K, Ca, Ti, V, Mn, Fe, and Ni) analyzed in this work.

\acknowledgments

We thank the anonymous referee for useful comments that helped improve the paper.
DS thanks Pascal Petit for the cordial host at CNRS-Toulouse where part of this project was developed.
KC and VS acknowledge that their work here is supported, in part, by the National Aeronautics and Space Administration under Grant 16-XRP16\_2-0004, issued through the Astrophysics Division of the Science Mission Directorate. 
D.A.G.H. and O.Z. acknowledge support from the State Research Agency  
(AEI) of the Spanish Ministry of Science, Innovation and Universities  
(MCIU) and the European Regional Development Fund (FEDER) under grant  
AYA2017-88254-P.
H. J. acknowledges support from the Crafoord Foundation, Stiftelsen Olle Engkvist Byggm\"astare, and Ruth och Nils-Erik Stenb\"acks stiftelse.

Funding for the Sloan Digital Sky Survey IV has been provided by the Alfred P. Sloan Foundation, the U.S. Department of Energy Office of Science, and the Participating Institutions. SDSS-IV acknowledges
support and resources from the Center for High-Performance Computing at the University of Utah. The SDSS web site is www.sdss.org.

SDSS-IV is managed by the Astrophysical Research consortium for the 
Participating Institutions of the SDSS Collaboration including the 
Brazilian Participation Group, the Carnegie Institution for Science, 
Carnegie Mellon University, the Chilean Participation Group, the French Participation Group, Harvard-Smithsonian Center for Astrophysics, 
Instituto de Astrof\'isica de Canarias, The Johns Hopkins University, 
Kavli Institute for the Physics and Mathematics of the Universe (IPMU)/University of Tokyo, Lawrence Berkeley National Laboratory, 
Leibniz Institut f\"ur Astrophysik Potsdam (AIP),  
Max-Planck-Institut f\"ur Astronomie (MPIA Heidelberg), 
Max-Planck-Institut f\"ur Astrophysik (MPA Garching), 
Max-Planck-Institut f\"ur Extraterrestrische Physik (MPE), 
National Astronomical Observatory of China, New Mexico State University, 
New York University, University of Notre Dame, 
Observat\'orio Nacional / MCTI, The Ohio State University, 
Pennsylvania State University, Shanghai Astronomical Observatory, 
United Kingdom Participation Group,
Universidad Nacional Aut\'onoma de M\'exico, University of Arizona, 
University of Colorado Boulder, University of Oxford, University of Portsmouth, 
University of Utah, University of Virginia, University of Washington, University of Wisconsin, 
Vanderbilt University, and Yale University.

\facilities{Sloan}


{}

\startlongtable
\begin{longrotatetable}
\begin{deluxetable}{lcccccccccccccccccc}
\tablenum{1}
\tabletypesize{\tiny}
\tablecaption{Stellar Properties}
\tablewidth{0pt}
\tablehead{
\colhead{2Mass ID} &
\colhead{RV} &
\colhead{$\sigma$(RV)} &
\colhead{PM (ra)} &
\colhead{PM (ra)} &
\colhead{PM (dec)} &
\colhead{PM (dec)} &
\colhead{Dist BJ18} &
\colhead{Dist ($\sigma$)} &
\colhead{SNR} &
\colhead{Prob} &
\colhead{Prob} &
\colhead{Prob} &
\colhead{Prob} &
\colhead{Prob} &
\colhead{$J$} & 
\colhead{$H$} & 
\colhead{$Ks$} \\
\colhead{} &
\colhead{(km s$^{-1}$)} &
\colhead{$\sigma$(km s$^{-1}$)} &
\colhead{($\mu_{\alpha}$ $\cos$($\delta$))} &
\colhead{$\sigma$} &
\colhead{($\mu_{\delta}$)} &
\colhead{$\sigma$} &
\colhead{Parsec} &
\colhead{Parsec} &
\colhead{} &
\colhead{G15} &
\colhead{Y08} &
\colhead{Z93} &
\colhead{G89} &
\colhead{S77} &
\colhead{} & 
\colhead{} & 
\colhead{}
}
\startdata
\textbf{Red Giants} \\
2M08492491+1144057	&	35.09	&	1.51	&	-11.06	&	0.07	&	-2.87	&	0.05	&	829.45	&	28.15	&	460	&	98	&	99	&	99	&	99	&	2	&	10.296	&	9.831	&	9.708\\ 
2M08503613+1143180	&	34.29	&	0.11	&	-11.06	&	0.07	&	-2.74	&	0.06	&	873.89	&	24.21	&	138	&	72	&	100	&	93	&	97	&	94	&	11.131	&	10.644	&	10.552\\ 
2M08504964+1135089	&	34.92	&	0.07	&	-10.96	&	0.08	&	-2.96	&	0.06	&	832.42	&	28.06	&	344	&	98	&	99	&	94	&	99	&	95	&	9.410	&	8.848	&	8.722\\ 
2M08511269+1152423	&	34.34	&	0.07	&	-10.95	&	0.06	&	-2.98	&	0.04	&	818.41	&	12.56	&	1445	&	98	&	99	&	99	&	96	&	95	&	8.650	&	8.122	&	7.976\\ 
2M08511704+1150464	&	33.58	&	0.06	&	-11.16	&	0.07	&	-3.32	&	0.05	&	829.68	&	28.19	&	371	&	98	&	99	&	77	&	97	&	95	&	9.284	&	8.712	&	8.606\\ 
2M08511897+1158110	&	34.01	&	0.10	&	-11.08	&	0.06	&	-3.09	&	0.04	&	847.08	&	23.29	&	384	&	98	&	100	&	94	&	98	&	51	&	10.587	&	10.095	&	10.012\\ 
2M08512156+1146061	&	34.87	&	0.06	&	-11.10	&	0.08	&	-2.66	&	0.05	&	834.37	&	31.65	&	314	&	97	&	98	&	91	&	99	&	95	&	9.602	&	9.085	&	8.947\\ 
2M08512618+1153520	&	34.16	&	0.04	&	-11.00	&	0.07	&	-2.88	&	0.05	&	842.43	&	14.79	&	982	&	97	&	97	&	77	&	99	&	95	&	8.619	&	8.113	&	7.960\\ 
2M08512898+1150330	&	33.46	&	0.04	&	-11.14	&	0.08	&	-3.22	&	0.05	&	812.29	&	18.46	&	481	&	98	&	100	&	94	&	98	&	95	&	8.566	&	8.072	&	7.958\\ 
2M08512990+1147168	&	36.28	&	0.01	&	-11.27	&	0.09	&	-3.73	&	0.05	&	795.23	&	34.60	&	884	&	98	&	99	&	0	&	96	&	96	&	7.314	&	6.681	&	6.489\\ 
2M08513577+1153347	&	34.05	&	0.11	&	-11.06	&	0.06	&	-2.93	&	0.04	&	801.49	&	20.07	&	205	&	98	&	93	&	99	&	72	&	95	&	10.522	&	10.023	&	9.941\\ 
2M08513938+1151456	&	33.98	&	0.11	&	-11.10	&	0.07	&	-3.12	&	0.04	&	834.61	&	26.61	&	469	&	98	&	100	&	95	&	99	&	93	&	10.383	&	9.889	&	9.795\\ 
2M08514234+1150076	&	34.27	&	0.05	&	-11.02	&	0.07	&	-2.80	&	0.05	&	805.95	&	23.64	&	271	&	98	&	0	&	99	&	99	&	96	&	9.829	&	9.339	&	9.187\\ 
2M08514388+1156425	&	32.94	&	0.05	&	-11.18	&	0.11	&	-3.16	&	0.07	&	844.60	&	22.86	&	505	&	95	&	100	&	99	&	98	&	91	&	8.618	&	8.114	&	7.996\\ 
2M08514507+1147459	&	32.97	&	0.04	&	-11.05	&	0.07	&	-3.03	&	0.04	&	839.67	&	26.03	&	281	&	97	&	99	&	2	&	92	&	92	&	9.684	&	9.183	&	9.045\\ 
2M08514883+1156511	&	34.35	&	0.05	&	-10.96	&	0.08	&	-3.26	&	0.05	&	858.95	&	35.02	&	135	&	97	&	99	&	99	&	99	&	94	&	11.256	&	10.779	&	10.705\\ 
2M08515611+1150147	&	34.68	&	0.04	&	-11.13	&	0.08	&	-3.89	&	0.05	&	843.67	&	28.77	&	133	&	98	&	99	&	99	&	98	&	95	&	11.197	&	10.726	&	10.634\\ 
2M08515952+1155049	&	34.39	&	0.05	&	-11.00	&	0.09	&	-3.10	&	0.06	&	868.56	&	21.45	&	543	&	98	&	99	&	88	&	91	&	90	&	8.597	&	8.084	&	7.959\\ 
2M08521097+1131491	&	33.82	&	0.03	&	-11.06	&	0.06	&	-2.76	&	0.04	&	822.72	&	23.87	&	672	&	98	&	100	&	92	&	98	&	96	&	8.921	&	8.388	&	8.252\\ 
2M08521656+1119380	&	33.82	&	0.03	&	-11.05	&	0.07	&	-2.88	&	0.05	&	808.61	&	30.73	&	1073	&	97	&	100	&	71	&	38	&	94	&	7.875	&	7.233	&	7.119\\ 
2M08521856+1144263	&	33.65	&	0.06	&	-11.13	&	0.07	&	-3.14	&	0.05	&	818.62	&	13.28	&	504	&	96	&	100	&	95	&	98	&	94	&	8.572	&	8.087	&	7.923\\ 
2M08522636+1141277	&	33.41	&	0.10	&	-10.77	&	0.08	&	-2.99	&	0.05	&	784.76	&	28.62	&	196	&	97	&	99	&	99	&	99	&	0	&	10.845	&	10.314	&	10.263\\ 
2M08525625+1148539	&	32.84	&	0.07	&	-11.02	&	0.08	&	-3.10	&	0.05	&	870.52	&	33.96	&	195	&	97	&	99	&	99	&	99	&	77	&	10.839	&	10.315	&	10.224\\ 
2M08534672+1123307	&	33.04	&	0.07	&	-11.24	&	0.08	&	-2.79	&	0.05	&	864.38	&	32.99	&	370	&	nan	&	nan	&	nan	&	nan	&	nan	&	10.225	&	9.730	&	9.624\\ 
2M08493465+1151256	&	33.98	&	0.08	&	-10.98	&	0.06	&	-2.92	&	0.04	&	904.59	&	nan	&	1369	&	98	&	98	&	99	&	91	&	96	&	7.203	&	6.546	&	6.394\\ 
2M08505816+1152223	&	34.03	&	0.11	&	-11.13	&	0.08	&	-2.86	&	0.05	&	884.25	&	34.01	&	287	&	98	&	99	&	99	&	96	&	91	&	11.197	&	10.707	&	10.626\\ 
2M08510723+1153019	&	32.99	&	30.95	&	-10.92	&	0.07	&	-2.41	&	0.05	&	903.98	&	21.87	&	604	&	11	&	100	&	99	&	97	&	71	&	11.175	&	10.771	&	10.695\\ 
2M08510839+1147121	&	33.52	&	0.18	&	-10.91	&	0.08	&	-2.93	&	0.06	&	888.63	&	28.60	&	171	&	98	&	98	&	99	&	99	&	93	&	10.691	&	10.195	&	10.112\\ 
2M08522003+1127362	&	33.94	&	0.05	&	-11.22	&	0.07	&	-2.91	&	0.04	&	893.78	&	28.66	&	260	&	98	&	99	&	93	&	91	&	89	&	10.839	&	10.383	&	10.253\\ 
\textbf{SubGiants} \\
2M08504994+1149127	&	33.83	&	0.10	&	-10.83	&	0.07	&	-3.27	&	0.05	&	809.79	&	28.26	&	110	&	98	&	100	&	94	&	97	&	93	&	11.372	&	10.960	&	10.890\\ 
2M08510325+1145473	&	35.11	&	0.21	&	-11.07	&	0.08	&	-2.91	&	0.06	&	829.30	&	27.10	&	103	&	61	&	96	&	54	&	99	&	95	&	11.491	&	11.220	&	11.187\\ 
2M08511564+1150561	&	34.01	&	0.03	&	-10.73	&	0.07	&	-2.78	&	0.05	&	785.07	&	21.58	&	209	&	86	&	100	&	99	&	98	&	94	&	11.485	&	11.094	&	11.013\\ 
2M08511670+1145293	&	35.48	&	0.21	&	-11.26	&	0.13	&	-2.41	&	0.09	&	843.16	&	51.73	&	140	&	nan	&	nan	&	nan	&	nan	&	nan	&	11.021	&	10.662	&	10.570\\ 
2M08512122+1145526	&	33.49	&	0.69	&	-11.74	&	0.09	&	-2.47	&	0.06	&	852.74	&	38.64	&	110	&	98	&	98	&	95	&	99	&	95	&	11.135	&	10.888	&	10.835\\ 
2M08512879+1151599	&	33.59	&	0.13	&	-10.91	&	0.07	&	-3.04	&	0.05	&	840.04	&	26.19	&	116	&	98	&	97	&	99	&	99	&	92	&	11.433	&	11.104	&	11.024\\ 
2M08512935+1145275	&	33.14	&	0.06	&	-10.74	&	0.07	&	-2.98	&	0.04	&	837.93	&	27.28	&	135	&	98	&	98	&	99	&	99	&	95	&	11.287	&	10.864	&	10.754\\ 
2M08513540+1157564	&	33.39	&	0.05	&	-11.10	&	0.07	&	-3.01	&	0.04	&	848.14	&	24.64	&	238	&	98	&	96	&	96	&	99	&	95	&	11.447	&	11.143	&	11.030\\ 
2M08513862+1220141	&	33.74	&	0.12	&	-10.95	&	0.08	&	-3.00	&	0.05	&	858.14	&	32.96	&	251	&	98	&	25	&	99	&	97	&	93	&	11.298	&	10.866	&	10.791\\ 
2M08514401+1146245	&	33.11	&	0.13	&	-11.10	&	0.07	&	-2.89	&	0.05	&	870.91	&	29.12	&	116	&	95	&	99	&	99	&	98	&	95	&	11.438	&	11.110	&	11.027\\ 
2M08514474+1146460	&	33.12	&	0.06	&	-11.06	&	0.07	&	-3.12	&	0.04	&	798.50	&	23.63	&	351	&	98	&	100	&	99	&	98	&	92	&	11.357	&	10.918	&	10.822\\ 
2M08514994+1149311	&	33.33	&	0.13	&	-11.35	&	0.07	&	-3.10	&	0.04	&	858.19	&	27.21	&	205	&	98	&	99	&	99	&	99	&	0	&	11.494	&	11.196	&	11.148\\ 
2M08515335+1148208	&	34.28	&	0.04	&	-11.44	&	0.07	&	-2.94	&	0.04	&	817.06	&	26.68	&	189	&	98	&	99	&	7.	&	nan	&	99	&	90	&	11.625	&	11.390\\
2M08521134+1145380	&	33.05	&	0.04	&	-10.98	&	0.07	&	-2.99	&	0.04	&	858.43	&	27.32	&	113	&	98	&	98	&	99	&	37	&	0	&	11.452	&	11.082	&	10.993\\ 
2M08503667+1148553	&	35.36	&	0.22	&	-11.43	&	0.06	&	-3.11	&	0.04	&	899.99	&	30.68	&	162	&	97	&	99	&	99	&	98	&	96	&	11.930	&	11.628	&	11.578\\ 
2M08505569+1152146	&	34.08	&	0.09	&	-11.01	&	0.18	&	-2.84	&	0.13	&	930.54	&	57.06	&	425	&	97	&	99	&	99	&	95	&	95	&	10.852	&	10.586	&	10.515\\ 
2M08510106+1150108	&	32.90	&	0.14	&	-10.79	&	0.09	&	-2.93	&	0.06	&	899.78	&	30.39	&	117	&	97	&	100	&	93	&	98	&	87	&	11.380	&	11.018	&	10.951\\ 
2M08510951+1141449	&	32.36	&	0.08	&	-10.33	&	0.07	&	-3.11	&	0.06	&	897.57	&	30.12	&	113	&	97	&	100	&	99	&	47	&	0	&	11.445	&	11.102	&	10.997\\ 
2M08511877+1151186	&	34.07	&	0.10	&	-10.98	&	0.07	&	-2.74	&	0.05	&	883.67	&	33.72	&	333	&	98	&	0.0	&	99	&	99	&	95	&	11.502	&	11.089	&	11.020\\ 
2M08515567+1217573	&	33.52	&	0.05	&	-10.99	&	0.08	&	-2.86	&	0.06	&	935.73	&	37.02	&	226	&	96	&	100	&	99	&	98	&	96	&	11.516	&	11.115	&	11.005\\ 
\textbf{turnoff} \\
2M08503392+1146272	&	33.78	&	0.14	&	-10.97	&	0.08	&	-3.05	&	0.06	&	869.53	&	32.54	&	241	&	98	&	99	&	99	&	99	&	96	&	11.824	&	11.596	&	11.517\\ 
2M08504079+1147462	&	34.59	&	0.06	&	-10.89	&	0.07	&	-3.08	&	0.05	&	847.32	&	28.41	&	170	&	98	&	99	&	86	&	98	&	93	&	11.793	&	11.540	&	11.498\\ 
2M08505177+1200247	&	33.72	&	0.25	&	-11.22	&	0.05	&	-2.85	&	0.05	&	867.93	&	21.00	&	146	&	75	&	99	&	16	&	94	&	95	&	12.377	&	12.106	&	12.051\\ 
2M08505702+1159158	&	33.96	&	0.22	&	-11.06	&	0.05	&	-3.72	&	0.03	&	840.73	&	21.18	&	178	&	98	&	100	&	99	&	98	&	94	&	12.003	&	11.726	&	11.673\\ 
2M08505762+1155147	&	33.07	&	0.24	&	-10.71	&	0.04	&	-2.85	&	0.03	&	870.75	&	17.25	&	148	&	98	&	100	&	94	&	98	&	93	&	12.294	&	12.038	&	11.973\\ 
2M08505903+1148576	&	33.67	&	0.44	&	-10.97	&	0.05	&	-2.73	&	0.03	&	867.89	&	21.19	&	118	&	94	&	100	&	99	&	99	&	96	&	12.386	&	12.206	&	12.094\\ 
2M08505973+1139524	&	33.21	&	0.19	&	-10.62	&	0.07	&	-2.74	&	0.05	&	831.18	&	28.29	&	104	&	98	&	100	&	99	&	99	&	96	&	12.025	&	11.735	&	11.703\\ 
2M08510969+1159096	&	33.98	&	7.27	&	-10.79	&	0.04	&	-2.93	&	0.03	&	857.36	&	14.81	&	154	&	98	&	99	&	99	&	72	&	94	&	12.658	&	12.348	&	12.298\\ 
2M08511576+1152587	&	35.82	&	0.08	&	-11.96	&	0.07	&	-2.03	&	0.05	&	844.46	&	29.65	&	162	&	98	&	99	&	99	&	99	&	64	&	11.728	&	11.453	&	11.391\\ 
2M08512240+1151291	&	33.44	&	0.16	&	-10.94	&	0.05	&	-2.96	&	0.04	&	853.50	&	20.51	&	132	&	98	&	99	&	94	&	84	&	79	&	12.195	&	11.952	&	11.862\\ 
2M08513710+1154599	&	34.85	&	0.05	&	-10.85	&	0.04	&	-2.95	&	0.03	&	858.09	&	19.44	&	127	&	96	&	99	&	99	&	0	&	0	&	12.096	&	11.819	&	11.763\\ 
2M08513806+1201243	&	32.14	&	0.12	&	-11.03	&	0.06	&	-3.40	&	0.04	&	839.80	&	28.69	&	144	&	98	&	99	&	99	&	99	&	69	&	11.844	&	11.551	&	11.495\\ 
2M08514122+1154290	&	33.61	&	0.21	&	-11.15	&	0.07	&	-3.06	&	0.05	&	820.44	&	25.67	&	213	&	83	&	98	&	99	&	81	&	94	&	11.703	&	11.466	&	11.397\\ 
2M08514475+1145012	&	34.89	&	0.20	&	-10.87	&	0.04	&	-2.81	&	0.03	&	855.48	&	18.41	&	136	&	98	&	97	&	99	&	99	&	95	&	12.288	&	12.039	&	11.969\\ 
2M08520741+1150221	&	34.19	&	0.16	&	-11.12	&	0.04	&	-2.95	&	0.03	&	864.00	&	18.90	&	202	&	98	&	99	&	99	&	77	&	83	&	12.097	&	11.823	&	11.806\\ 
\textbf{main-sequence} \\
2M08502805+1154505	&	34.95	&	0.20	&	-10.55	&	0.04	&	-2.41	&	0.03	&	860.00	&	19.59	&	122	&	98	&	100	&	99	&	99	&	93	&	12.968	&	12.665	&	12.563\\ 
2M08511229+1154230	&	35.13	&	0.26	&	-10.81	&	0.05	&	-2.87	&	0.04	&	850.83	&	21.65	&	118	&	98	&	97	&	99	&	98	&	91	&	12.986	&	12.708	&	12.623\\ 
2M08512314+1154049	&	33.62	&	0.33	&	-10.83	&	0.05	&	-2.76	&	0.03	&	846.83	&	20.20	&	119	&	98	&	98	&	99	&	93	&	94	&	13.017	&	12.741	&	12.681\\ 
2M08512604+1149555	&	32.89	&	0.30	&	-11.76	&	0.06	&	-3.28	&	0.04	&	853.19	&	25.88	&	137	&	98	&	97	&	99	&	98	&	92	&	13.344	&	12.987	&	12.897\\ 
2M08512996+1151090	&	34.82	&	0.23	&	-11.07	&	0.05	&	-3.08	&	0.03	&	855.38	&	22.12	&	192	&	98	&	99	&	89	&	99	&	93	&	12.926	&	12.630	&	12.599\\ 
2M08513119+1153179	&	34.17	&	0.30	&	-10.82	&	0.04	&	-2.98	&	0.03	&	858.47	&	19.64	&	156	&	98	&	99	&	99	&	97	&	95	&	12.603	&	12.327	&	12.267\\ 
2M08513701+1136516	&	33.18	&	0.67	&	-10.80	&	0.06	&	-3.31	&	0.04	&	848.99	&	24.24	&	100	&	98	&	100	&	92	&	98	&	94	&	13.341	&	12.932	&	12.829\\ 
2M08514189+1149376	&	35.88	&	0.28	&	-11.02	&	0.06	&	-2.92	&	0.04	&	862.75	&	25.07	&	112	&	98	&	98	&	99	&	98	&	95	&	13.626	&	13.262	&	13.189\\ 
2M08514742+1147096	&	31.44	&	6.03	&	-11.11	&	0.05	&	-3.09	&	0.03	&	840.28	&	19.47	&	113	&	98	&	99	&	95	&	99	&	96	&	12.880	&	12.496	&	12.372\\ 
2M08521649+1147382	&	33.91	&	0.27	&	-11.01	&	0.07	&	-2.74	&	0.04	&	814.76	&	28.03	&	120	&	98	&	99	&	99	&	99	&	77	&	13.558	&	13.221	&	13.157\\ 
2M08505439+1156290	&	33.73	&	0.11	&	-10.74	&	0.07	&	-3.19	&	0.05	&	919.56	&	34.20	&	270	&	98	&	98	&	99	&	99	&	95	&	11.706	&	11.435	&	11.372\\ 
2M08510076+1153115	&	34.05	&	0.28	&	-10.76	&	0.06	&	-2.93	&	0.05	&	914.22	&	28.97	&	119	&	96	&	100	&	96	&	97	&	93	&	13.474	&	13.157	&	13.105\\ 
2M08511176+1150018	&	33.53	&	0.29	&	-11.02	&	0.08	&	-3.20	&	0.05	&	899.09	&	29.78	&	127	&	3	&	100	&	96	&	98	&	92	&	13.665	&	13.120	&	13.031\\ 
2M08512080+1145024	&	33.77	&	0.11	&	-10.51	&	0.06	&	-3.81	&	0.04	&	889.07	&	28.43	&	109	&	nan	&	89	&	97	&	95	&	99	&	11.928	&	11.679	&	11.603\\ 
2M08512742+1153265	&	34.28	&	0.18	&	-10.87	&	0.06	&	-3.09	&	0.04	&	948.69	&	34.44	&	273	&	95	&	96	&	99	&	98	&	95	&	11.667	&	11.382	&	11.342\\ 
2M08512788+1155409	&	36.06	&	0.18	&	-11.13	&	0.04	&	-2.40	&	0.03	&	893.36	&	18.68	&	129	&	98	&	100	&	99	&	99	&	95	&	12.168	&	11.831	&	11.813\\ 
2M08513012+1143498	&	33.53	&	0.23	&	-11.11	&	0.09	&	-3.09	&	0.06	&	887.79	&	38.35	&	102	&	98	&	1	&	99	&	99	&	99	&	12.011	&	11.761	&	11.694\\ 
2M08513455+1149068	&	33.53	&	0.41	&	-11.05	&	0.08	&	-3.20	&	0.05	&	887.52	&	27.47	&	104	&	98	&	91	&	96	&	99	&	95	&	13.717	&	13.229	&	13.121\\ 
2M08521868+1143246	&	32.73	&	0.18	&	-10.97	&	0.07	&	-2.86	&	0.04	&	877.95	&	32.18	&	143	&	98	&	99	&	88	&	99	&	75	&	11.590	&	11.352	&	11.259\\ 
2M08512643+1143506	&	33.45	&	0.26	&	-11.37	&	0.11	&	-2.69	&	0.08	&	835.57	&	38.90	&	121	&	98	&	100	&	96	&	99	&	95	&	11.020	&	11.011	&	10.993\\ 
2M08513259+1148520	&	33.74	&	0.27	&	-11.30	&	0.08	&	-3.11	&	0.05	&	791.70	&	28.99	&	146	&	96	&	98	&	99	&	87	&	0	&	10.645	&	10.541	&	10.526\\ 
\hline
Excluded Sample due\\
to low SNR ($<$) 100\\
\textbf{SubGiant} \\
2M08503438+1139566	&	33.77	&	0.21	&	-10.79	&	0.07	&	-2.94	&	0.05	&	850.43	&	27.79	&	99	&	98	&	99	&	91	&	98	&	96	&	11.513	&	11.244	&	11.177\\ 
2M08504198+1136525	&	34.46	&	0.10	&	-11.12	&	0.07	&	-3.10	&	0.05	&	852.64	&	27.51	&	91	&	99	&	99	&	99	&	99	&	0	&	11.410	&	11.062	&	10.998\\ 
2M08510811+1201065	&	33.83	&	0.23	&	-11.70	&	0.04	&	-3.02	&	0.03	&	875.47	&	20.06	&	99	&	98	&	96	&	99	&	99	&	95	&	12.469	&	12.159	&	12.073\\ 
2M08511826+1150196	&	34.28	&	4.21	&	-10.89	&	0.06	&	-2.59	&	0.04	&	863.61	&	27.89	&	86	&	98	&	99	&	94	&	98	&	96	&	13.042	&	12.680	&	12.592\\ 
2M08520356+1141238	&	34.15	&	0.07	&	-10.82	&	0.07	&	-2.76	&	0.04	&	852.79	&	27.94	&	99	&	98	&	99	&	99	&	99	&	89	&	11.634	&	11.365	&	11.306\\ 
\textbf{main-sequence} \\
2M08502833+1142097	&	33.75	&	0.39	&	-10.85	&	0.07	&	-2.83	&	0.05	&	832.76	&	25.30	&	64	&	98	&	97	&	99	&	98	&	88	&	11.899	&	11.654	&	11.587\\ 
2M08503788+1252295	&	32.37	&	0.30	&	-11.83	&	0.07	&	-3.44	&	0.05	&	801.51	&	25.81	&	99	&	nan	&	nan	&	nan	&	nan	&	nan	&	13.662	&	13.239	&	13.139\\ 
2M08505334+1143399	&	32.72	&	0.33	&	-10.89	&	0.05	&	-3.92	&	0.04	&	841.16	&	21.35	&	89	&	98	&	100	&	99	&	98	&	94	&	13.058	&	12.746	&	12.628\\ 
2M08505923+1146129	&	31.98	&	1.78	&	-10.90	&	0.05	&	-2.84	&	0.03	&	810.49	&	18.26	&	55	&	98	&	100	&	99	&	98	&	96	&	12.271	&	11.998	&	11.934\\ 
2M08512386+1138521	&	34.61	&	0.35	&	-11.04	&	0.06	&	-2.81	&	0.05	&	849.79	&	23.97	&	89	&	98	&	99	&	99	&	80	&	95	&	13.313	&	12.952	&	12.942\\ 
2M08513215+1136126	&	34.34	&	0.41	&	-11.22	&	0.04	&	-2.83	&	0.03	&	869.31	&	17.26	&	52	&	94	&	96	&	99	&	99	&	89	&	12.207	&	11.965	&	11.910\\ 
2M08513444+1137574	&	34.02	&	0.12	&	-10.78	&	0.04	&	-2.66	&	0.03	&	829.37	&	17.97	&	58	&	98	&	97	&	99	&	67	&	1	&	12.102	&	11.864	&	11.778\\ 
2M08514375+1145148	&	32.40	&	0.19	&	-11.22	&	0.06	&	-2.94	&	0.04	&	848.12	&	24.69	&	55	&	98	&	97	&	99	&	99	&	95	&	12.027	&	11.805	&	11.729\\ 
2M08514465+1141510	&	32.95	&	0.27	&	-11.33	&	0.04	&	-2.95	&	0.03	&	860.46	&	17.58	&	67	&	97	&	99	&	99	&	98	&	95	&	12.120	&	11.887	&	11.802\\ 
2M08515290+1146358	&	34.00	&	0.53	&	-11.09	&	0.09	&	-2.78	&	0.05	&	865.42	&	23.43	&	97	&	98	&	94	&	99	&	99	&	94	&	13.961	&	13.429	&	13.282\\ 
2M08521664+1142300	&	32.23	&	3.99	&	-10.94	&	0.05	&	-2.95	&	0.03	&	805.89	&	17.43	&	70	&	98	&	96	&	99	&	97	&	92	&	12.403	&	12.144	&	12.104\\ 
2M08504511+1136023	&	31.15	&	13.33	&	-10.81	&	0.08	&	-2.70	&	0.07	&	842.68	&	17.27	&	76	&	97	&	99	&	99	&	98	&	71	&	13.800	&	13.210	&	13.123\\ 
2M08510131+1141587	&	32.07	&	10.61	&	-11.04	&	0.04	&	-2.81	&	0.03	&	879.70	&	20.72	&	80	&	96	&	100	&	99	&	98	&	95	&	12.420	&	12.167	&	12.075\\ 
2M08510156+1147501	&	32.93	&	0.23	&	-10.89	&	0.05	&	-3.60	&	0.04	&	876.51	&	23.19	&	57	&	98	&	100	&	93	&	98	&	96	&	12.371	&	12.067	&	11.991\\ 
2M08511229+1146212	&	31.51	&	0.28	&	-10.93	&	0.05	&	-3.04	&	0.04	&	919.79	&	20.28	&	73	&	98	&	99	&	93	&	91	&	91	&	12.060	&	11.751	&	11.704\\ 
2M08511810+1142547	&	33.88	&	0.11	&	-10.96	&	0.05	&	-2.89	&	0.04	&	923.46	&	25.50	&	97	&	99	&	99	&	0	&	84	&	94	&	12.186	&	11.879	&	11.844\\ 
2M08512033+1145523	&	33.66	&	0.29	&	-10.89	&	0.06	&	-2.96	&	0.04	&	876.62	&	24.68	&	76	&	98	&	84	&	99	&	99	&	94	&	12.061	&	11.822	&	11.767\\ 
2M08512176+1144050	&	32.79	&	0.46	&	-11.28	&	0.05	&	-3.12	&	0.04	&	881.45	&	24.05	&	60	&	96	&	99	&	96	&	97	&	96	&	12.907	&	12.547	&	12.498\\ 
2M08512467+1143061	&	32.01	&	0.32	&	-10.88	&	0.07	&	-2.18	&	0.05	&	926.54	&	36.75	&	82	&	98	&	99	&	99	&	83	&	94	&	13.258	&	12.863	&	12.806\\ 
2M08513424+1145535	&	34.19	&	0.46	&	-10.83	&	0.07	&	-2.81	&	0.05	&	941.80	&	35.23	&	77	&	98	&	4	&	99	&	99	&	95	&	13.374	&	12.976	&	12.852\\
\tablewidth{0pt}	
\enddata
\tablenotetext{}{Proper motions and distances from Gaia DR2}
\end{deluxetable}
\end{longrotatetable}

\begin{longrotatetable}
\begin{deluxetable}{lcccccccc}
\tablenum{2}
\tabletypesize{\tiny}
\tablecaption{Stellar Parameters}
\tablewidth{0pt}
\tablehead{
\colhead{2Mass ID} &
\colhead{$T_{\rm eff}$ (K)} &
\colhead{$T_{\rm eff}$ (K)} &
\colhead{$T_{\rm eff}$ (K)} &
\colhead{log $g$ (cm s$^{-2}$)} &
\colhead{log $g$ (cm s$^{-2}$)} &
\colhead{log $g$ (cm s$^{-2}$)} &
\colhead{Mass ($M_{\odot}$)} &
\colhead{$\xi$ (km s$^{-1}$)} \\
\colhead{} &
\colhead{ASPCAP raw} &
\colhead{ASPCAP calib} &
\colhead{GB09} &
\colhead{ASPCAP raw} &
\colhead{ASPCAP calib} &
\colhead{Physical} &
\colhead{Mass Isocronae} &
\colhead{ASPCAP raw}
}
\startdata
\textbf{Red Giants}\\
2M08492491+1144057	&	4848.2	&	4893.5	&	4899.0	&	3.31	&	3.17	&	3.15	&	1.32	&	1.14\\ 
2M08503613+1143180	&	4973.6	&	5023.9	&	5019.9	&	3.53	&	3.41	&	3.51	&	1.31	&	1.24\\ 
2M08504964+1135089	&	4727.8	&	4774.7	&	4710.9	&	2.96	&	2.80	&	2.77	&	1.33	&	1.33\\ 
2M08511269+1152423	&	4758.1	&	4805.4	&	4702.4	&	2.83	&	2.49	&	2.50	&	1.33	&	1.45\\ 
2M08511704+1150464	&	4707.8	&	4757.9	&	4764.2	&	2.87	&	2.71	&	2.71	&	1.33	&	1.42\\ 
2M08511897+1158110	&	4907.5	&	4956.5	&	4909.9	&	3.35	&	3.22	&	3.30	&	1.32	&	1.11\\ 
2M08512156+1146061	&	4731.2	&	4776.9	&	4748.2	&	3.01	&	2.85	&	2.86	&	1.33	&	1.35\\ 
2M08512618+1153520	&	4750.6	&	4798.5	&	4714.2	&	2.81	&	2.48	&	2.49	&	1.33	&	1.40\\ 
2M08512898+1150330	&	4693.6	&	4741.1	&	4691.2	&	2.79	&	2.46	&	2.47	&	1.34	&	1.43\\ 
2M08512990+1147168	&	4247.5	&	4302.4	&	4274.0	&	1.95	&	1.68	&	1.82	&	1.34	&	1.47\\ 
2M08513577+1153347	&	4911.4	&	4959.5	&	4882.8	&	3.37	&	3.24	&	3.29	&	1.32	&	1.19\\ 
2M08513938+1151456	&	4878.3	&	4927.4	&	4871.1	&	3.32	&	3.20	&	3.22	&	1.32	&	1.20\\ 
2M08514234+1150076	&	4778.7	&	4825.8	&	4803.5	&	3.13	&	2.99	&	2.97	&	1.33	&	1.27\\ 
2M08514388+1156425	&	4747.5	&	4795.8	&	4711.9	&	2.76	&	2.44	&	2.50	&	1.33	&	1.64\\ 
2M08514507+1147459	&	4765.1	&	4812.5	&	4799.6	&	3.08	&	2.93	&	2.88	&	1.33	&	1.35\\ 
2M08514883+1156511	&	4976.0	&	5027.4	&	5028.8	&	3.56	&	3.44	&	3.57	&	1.31	&	1.25\\ 
2M08515611+1150147	&	4950.8	&	4994.9	&	4927.9	&	3.70	&	3.56	&	3.56	&	1.31	&	0.92\\ 
2M08515952+1155049	&	4740.0	&	4789.0	&	4708.4	&	2.76	&	2.45	&	2.49	&	1.33	&	1.51\\ 
2M08521097+1131491	&	4602.3	&	4649.1	&	4633.6	&	2.75	&	2.56	&	2.55	&	1.33	&	1.38\\ 
2M08521656+1119380	&	4345.3	&	4394.8	&	4406.8	&	2.32	&	2.09	&	2.08	&	1.35	&	1.39\\ 
2M08521856+1144263	&	4702.7	&	4750.1	&	4737.3	&	2.75	&	2.44	&	2.44	&	1.35	&	1.49\\ 
2M08522636+1141277	&	4912.3	&	4962.6	&	4957.1	&	3.39	&	3.26	&	3.37	&	1.31	&	1.30\\ 
2M08525625+1148539	&	4899.5	&	4944.1	&	4874.3	&	3.42	&	3.28	&	3.38	&	1.31	&	1.29\\ 
2M08534672+1123307	&	4850.6	&	4899.3	&	4880.3	&	3.23	&	3.11	&	3.05	&	1.32	&	1.10\\ 
2M08493465+1151256	&	4190.5	&	-9999.0	&	4347.6	&	1.62	&	-9999.	&	1.78	&	1.34	&	0.53\\ 
2M08505816+1152223	&	4983.5	&	5030.7	&	5021.5	&	3.59	&	3.46	&	3.46	&	1.31	&	1.05\\ 
2M08510723+1153019	&	4661.0	&	-9999.0	&	5335.0	&	3.44	&	-9999.	&	2.62	&	1.33	&	0.72\\ 
2M08510839+1147121	&	4901.4	&	4948.9	&	4995.6	&	3.40	&	3.27	&	3.22	&	1.32	&	1.25\\ 
2M08522003+1127362	&	4932.4	&	4980.3	&	4973.2	&	3.42	&	3.29	&	3.28	&	1.32	&	1.20\\ 
\textbf{Subgiants}\\
2M08504994+1149127	&	5160.9	&	5213.0	&	5196.8	&	3.74	&	3.62	&	3.65	&	1.30	&	0.96\\ 
2M08510325+1145473	&	5884.4	&	5928.1	&	5932.2	&	4.17	&	-9999.	&	3.84	&	1.30	&	0.77\\ 
2M08511564+1150561	&	5282.5	&	5331.6	&	5271.0	&	3.81	&	-9999.	&	3.74	&	1.30	&	1.04\\ 
2M08511670+1145293	&	5280.6	&	5335.8	&	5312.0	&	3.86	&	-9999.	&	3.72	&	1.24	&	0.87\\ 
2M08512122+1145526	&	5926.6	&	5971.3	&	6019.4	&	4.21	&	-9999.	&	3.67	&	1.31	&	0.75\\ 
2M08512879+1151599	&	5617.4	&	5673.6	&	5624.3	&	3.91	&	-9999.	&	3.82	&	1.29	&	0.78\\ 
2M08512935+1145275	&	5019.9	&	5069.7	&	5061.7	&	3.61	&	3.49	&	3.60	&	1.31	&	1.16\\ 
2M08513540+1157564	&	5446.4	&	5497.9	&	5484.9	&	3.86	&	-9999.	&	3.74	&	1.29	&	0.65\\ 
2M08513862+1220141	&	5062.5	&	5112.0	&	4995.6	&	3.70	&	3.58	&	3.66	&	1.30	&	0.94\\ 
2M08514401+1146245	&	5432.7	&	5483.9	&	5507.6	&	3.85	&	-9999.	&	3.73	&	1.30	&	0.89\\ 
2M08514474+1146460	&	5058.6	&	5109.2	&	5065.6	&	3.69	&	3.57	&	3.64	&	1.30	&	0.97\\ 
2M08514994+1149311	&	6003.3	&	6048.3	&	5887.6	&	4.11	&	-9999.	&	3.85	&	1.28	&	0.83\\ 
2M08515335+1148208	&	6069.2	&	6114.7	&	5991.7	&	4.10	&	-9999.	&	3.89	&	1.27	&	1.06\\ 
2M08521134+1145380	&	5293.2	&	5343.2	&	5326.3	&	3.81	&	-9999.	&	3.71	&	1.30	&	0.92\\ 
2M08503667+1148553	&	5689.4	&	5746.5	&	5874.4	&	3.92	&	-9999.	&	3.82	&	1.29	&	0.68\\ 
2M08505569+1152146	&	5910.2	&	5954.9	&	5943.0	&	4.09	&	-9999.	&	3.85	&	1.28	&	0.67\\ 
2M08510106+1150108	&	5428.1	&	5480.6	&	5540.4	&	3.92	&	-9999.	&	3.78	&	1.29	&	0.82\\ 
2M08510951+1141449	&	5445.3	&	5492.3	&	5462.4	&	3.89	&	-9999.	&	3.77	&	1.29	&	0.84\\ 
2M08511877+1151186	&	5231.5	&	5277.7	&	5326.9	&	3.85	&	-9999.	&	3.73	&	1.30	&	0.81\\ 
2M08515567+1217573	&	5213.3	&	5264.6	&	5215.4	&	3.78	&	3.66	&	3.73	&	1.30	&	0.91\\ 
\textbf{turnoff}\\
2M08503392+1146272	&	6235.1	&	6279.1	&	6165.5	&	4.31	&	-9999.	&	4.02	&	1.25	&	0.73\\ 
2M08504079+1147462	&	6228.8	&	6274.8	&	6156.4	&	4.22	&	-9999.	&	3.97	&	1.25	&	0.72\\ 
2M08505177+1200247	&	6009.0	&	6053.3	&	6042.0	&	4.22	&	-9999.	&	4.15	&	1.21	&	0.62\\ 
2M08505702+1159158	&	6024.5	&	6069.9	&	6040.1	&	4.20	&	-9999.	&	4.03	&	1.27	&	0.55\\ 
2M08505762+1155147	&	6151.3	&	6196.3	&	6044.2	&	4.32	&	-9999.	&	4.16	&	1.21	&	0.64\\ 
2M08505903+1148576	&	5996.1	&	-9999.0	&	6090.9	&	4.25	&	-9999.	&	4.17	&	1.21	&	0.66\\ 
2M08505973+1139524	&	6061.8	&	6107.0	&	5968.9	&	4.34	&	-9999.	&	4.08	&	1.25	&	0.64\\ 
2M08510969+1159096	&	6026.4	&	6073.3	&	5957.1	&	4.18	&	-9999.	&	4.08	&	1.25	&	0.63\\ 
2M08511576+1152587	&	6093.1	&	6137.2	&	5960.3	&	4.18	&	-9999.	&	3.97	&	1.27	&	1.24\\ 
2M08512240+1151291	&	6009.8	&	6056.5	&	6056.2	&	4.20	&	-9999.	&	4.08	&	1.23	&	0.85\\ 
2M08513710+1154599	&	6110.2	&	6156.0	&	6052.5	&	4.29	&	-9999.	&	4.07	&	1.23	&	0.63\\ 
2M08513806+1201243	&	5882.2	&	5926.9	&	5845.0	&	4.16	&	-9999.	&	3.95	&	1.28	&	0.59\\ 
2M08514122+1154290	&	6118.1	&	6162.7	&	6008.6	&	4.19	&	-9999.	&	3.95	&	1.28	&	0.74\\ 
2M08514475+1145012	&	6040.5	&	6086.9	&	6136.3	&	4.36	&	-9999.	&	4.14	&	1.23	&	0.68\\ 
2M08520741+1150221	&	6043.8	&	6087.8	&	6057.0	&	4.29	&	-9999.	&	4.08	&	1.25	&	0.66\\ 
\textbf{main-sequence}\\
2M08502805+1154505	&	5759.3	&	5806.5	&	5755.8	&	4.17	&	-9999.	&	4.33	&	1.14	&	0.61\\ 
2M08511229+1154230	&	5848.9	&	5892.7	&	5885.6	&	4.35	&	-9999.	&	4.34	&	1.12	&	0.68\\ 
2M08512314+1154049	&	5802.4	&	5847.1	&	5886.3	&	4.27	&	-9999.	&	4.34	&	1.11	&	0.73\\ 
2M08512604+1149555	&	5310.1	&	5358.8	&	5472.1	&	4.08	&	-9999.	&	4.33	&	1.00	&	0.61\\ 
2M08512996+1151090	&	5900.8	&	5945.5	&	5925.8	&	4.34	&	-9999.	&	4.34	&	1.11	&	0.64\\ 
2M08513119+1153179	&	6062.9	&	6108.2	&	6021.5	&	4.30	&	-9999.	&	4.24	&	1.17	&	0.68\\ 
2M08513701+1136516	&	5201.7	&	5244.9	&	5211.6	&	4.52	&	-9999.	&	4.65	&	0.81	&	0.86\\ 
2M08514189+1149376	&	5481.9	&	5525.2	&	5595.9	&	4.31	&	-9999.	&	4.48	&	1.03	&	0.90\\ 
2M08514742+1147096	&	5199.1	&	5245.7	&	5226.8	&	4.24	&	-9999.	&	4.57	&	0.85	&	0.60\\ 
2M08521649+1147382	&	5467.4	&	5512.1	&	5573.0	&	4.33	&	-9999.	&	4.49	&	1.01	&	0.72\\ 
2M08505439+1156290	&	6163.0	&	6207.7	&	5956.9	&	4.29	&	-9999.	&	4.33	&	1.10	&	0.63\\ 
2M08510076+1153115	&	5564.3	&	5609.2	&	5798.0	&	4.25	&	-9999.	&	4.50	&	0.93	&	0.60\\ 
2M08511176+1150018	&	4857.4	&	4902.6	&	4779.6	&	4.20	&	-9999.	&	4.67	&	0.75	&	0.57\\ 
2M08512080+1145024	&	5799.6	&	5845.3	&	5998.8	&	4.28	&	-9999.	&	4.40	&	0.98	&	0.67\\ 
2M08512742+1153265	&	6169.5	&	6214.1	&	5998.8	&	4.31	&	-9999.	&	4.31	&	1.11	&	0.73\\ 
2M08512788+1155409	&	5834.3	&	5878.5	&	5860.8	&	4.30	&	-9999.	&	4.48	&	0.98	&	0.61\\ 
2M08513012+1143498	&	5866.0	&	5913.1	&	6036.4	&	4.36	&	-9999.	&	4.45	&	1.00	&	0.62\\ 
2M08513455+1149068	&	4922.6	&	4966.8	&	4908.2	&	4.45	&	-9999.	&	4.69	&	0.75	&	0.86\\ 
2M08521868+1143246	&	5960.3	&	6006.6	&	5970.8	&	4.19	&	-9999.	&	4.42	&	1.03	&	0.62\\ 
2M08512643+1143506	&	7985.2	&	-9999.0	&	7686.9	&	4.67	&	-9999.	&	4.53	&	1.21	&	1.55\\ 
2M08513259+1148520	&	7599.0	&	7647.2	&	7095.6	&	4.38	&	-9999.	&	4.44	&	1.21	&	2.82\\ 
\hline
Excluded Sample due\\
to low SNR ($<$) 100\\
\textbf{Subgiant} \\
2M08503438+1139566	&	5955.8	&	5998.6	&	5947.6	&	4.11	&	-9999.	&	3.87	&	1.28	&	0.67\\ 
2M08504198+1136525	&	5597.0	&	5647.9	&	5611.6	&	3.92	&	-9999.	&	3.79	&	1.29	&	0.94\\ 
2M08510811+1201065	&	5625.6	&	5674.2	&	5679.9	&	4.02	&	-9999.	&	3.80	&	1.29	&	0.59\\ 
2M08511826+1150196	&	5458.1	&	5508.6	&	5454.3	&	4.01	&	-9999.	&	3.77	&	1.29	&	0.58\\ 
2M08520356+1141238	&	5954.6	&	6000.4	&	5984.7	&	4.07	&	-9999.	&	3.89	&	1.28	&	0.70\\ 
\textbf{main-sequence}\\
2M08502833+1142097	&	6164.8	&	6210.0	&	6060.1	&	4.28	&	-9999.	&	4.33	&	1.10	&	0.86\\ 
2M08503788+1252295	&	5362.5	&	5405.2	&	5381.2	&	4.39	&	-9999.	&	4.68	&	0.87	&	1.11\\ 
2M08505334+1143399	&	5194.7	&	5242.1	&	5536.2	&	4.22	&	-9999.	&	4.56	&	0.85	&	0.69\\ 
2M08505923+1146129	&	5567.1	&	5613.8	&	5943.0	&	4.21	&	-9999.	&	4.51	&	0.93	&	0.71\\ 
2M08512386+1138521	&	5614.1	&	5659.0	&	5789.1	&	4.47	&	-9999.	&	4.52	&	0.93	&	0.70\\ 
2M08513215+1136126	&	5935.4	&	5981.2	&	6132.1	&	4.33	&	-9999.	&	4.42	&	1.03	&	0.73\\ 
2M08513444+1137574	&	6000.8	&	6045.6	&	6003.4	&	4.24	&	-9999.	&	4.40	&	1.04	&	0.63\\ 
2M08514375+1145148	&	6056.9	&	6104.1	&	6127.2	&	4.30	&	-9999.	&	4.38	&	1.06	&	0.66\\ 
2M08514465+1141510	&	6091.1	&	6136.1	&	6031.7	&	4.39	&	-9999.	&	4.36	&	1.07	&	0.72\\ 
2M08515290+1146358	&	4624.6	&	4670.1	&	4633.9	&	4.34	&	-9999.	&	4.71	&	0.70	&	1.24\\ 
2M08521664+1142300	&	5985.2	&	6033.1	&	6122.4	&	4.28	&	-9999.	&	4.36	&	1.04	&	0.58\\ 
2M08504511+1136023	&	4547.0	&	4593.2	&	4640.2	&	4.17	&	-9999.	&	4.55	&	0.75	&	0.57\\ 
2M08510131+1141587	&	5904.2	&	5953.5	&	5906.2	&	4.24	&	-9999.	&	4.45	&	1.01	&	0.70\\ 
2M08510156+1147501	&	5642.9	&	5688.3	&	5749.4	&	4.17	&	-9999.	&	4.49	&	0.95	&	0.58\\ 
2M08511229+1146212	&	5806.2	&	5851.6	&	5856.3	&	4.25	&	-9999.	&	4.47	&	0.98	&	0.64\\ 
2M08511810+1142547	&	5895.8	&	5940.7	&	5920.0	&	4.27	&	-9999.	&	4.45	&	1.01	&	0.70\\ 
2M08512033+1145523	&	6063.8	&	6108.9	&	6146.6	&	4.32	&	-9999.	&	4.39	&	1.06	&	0.68\\ 
2M08512176+1144050	&	5360.8	&	5406.7	&	5624.3	&	4.26	&	-9999.	&	4.60	&	0.86	&	0.64\\ 
2M08512467+1143061	&	5219.1	&	5264.6	&	5446.2	&	4.39	&	-9999.	&	4.57	&	0.85	&	0.78\\ 
2M08513424+1145535	&	5193.5	&	5235.5	&	5174.0	&	4.62	&	-9999.	&	4.56	&	0.85	&	0.57\\ 
\tablewidth{0pt}	
\enddata
\end{deluxetable}
\end{longrotatetable}

\begin{longrotatetable}
\begin{deluxetable}{lrrrrrrrrrrrrrrrrr}
\tablenum{3}
\tabletypesize{\tiny}
\tablecaption{Stellar Abundances}
\tablewidth{0pt}
\tablehead{
\colhead{2Mass ID} &
\colhead{$[$Fe/H$]$} &
\colhead{$[$C/H$]$} &
\colhead{$[$N/H$]$} &
\colhead{$[$O/H$]$} &
\colhead{$[$Na/H$]$} &
\colhead{$[$Mg/H$]$} &
\colhead{$[$Al/H$]$} &
\colhead{$[$Si/H$]$} &
\colhead{$[$K/H$]$} &
\colhead{$[$Ca/H$]$} &
\colhead{$[$Ti/H$]$} &
\colhead{$[$V/H$]$} &
\colhead{$[$Cr/H$]$} &
\colhead{$[$Mn/H$]$} &
\colhead{$[$Ni/H$]$}
}
\startdata
\textbf{Red Giant} \\
2M08492491+1144057	&	0.13	&	-0.20	&	0.35	&	0.09	&	0.33	&	0.17	&	0.42	&	0.24	&	0.03	&	0.14	&	0.22	&	0.20	&	0.01	&	0.08	&	0.11\\ 
2M08503613+1143180	&	0.02	&	-0.07	&	0.14	&	0.04	&	0.08	&	0.02	&	0.18	&	0.13	&	-0.04	&	-0.01	&	-0.07	&	0.19	&	-0.01	&	-0.03	&	0.02\\ 
2M08504964+1135089	&	0.07	&	-0.16	&	0.43	&	0.09	&	0.32	&	0.11	&	0.32	&	0.21	&	0.00	&	0.06	&	0.04	&	0.28	&	-0.04	&	0.07	&	0.07\\ 
2M08511269+1152423	&	0.07	&	-0.16	&	0.29	&	0.11	&	0.32	&	0.13	&	0.47	&	0.26	&	-0.01	&	0.10	&	0.12	&	0.17	&	-0.04	&	0.08	&	0.03\\ 
2M08511704+1150464	&	0.07	&	-0.20	&	0.48	&	0.13	&	0.28	&	0.13	&	0.27	&	0.21	&	-0.03	&	0.06	&	0.08	&	-0.00	&	0.01	&	0.03	&	0.05\\ 
2M08511897+1158110	&	0.06	&	-0.20	&	0.38	&	0.13	&	0.14	&	0.07	&	0.26	&	0.19	&	-0.04	&	0.05	&	-0.02	&	0.09	&	-0.01	&	0.05	&	0.06\\ 
2M08512156+1146061	&	0.10	&	-0.15	&	0.35	&	0.12	&	0.31	&	0.14	&	0.34	&	0.24	&	0.01	&	0.09	&	0.06	&	0.26	&	0.01	&	0.08	&	0.09\\ 
2M08512618+1153520	&	0.06	&	-0.19	&	0.33	&	0.10	&	0.37	&	0.11	&	0.41	&	0.25	&	-0.05	&	0.06	&	0.08	&	0.14	&	-0.08	&	0.05	&	0.02\\ 
2M08512898+1150330	&	0.04	&	-0.08	&	0.21	&	0.09	&	0.35	&	0.10	&	0.34	&	0.24	&	-0.04	&	0.06	&	0.01	&	0.12	&	-0.08	&	0.06	&	0.03\\ 
2M08512990+1147168	&	-0.05	&	-0.14	&	0.39	&	0.05	&	0.35	&	0.07	&	0.31	&	0.11	&	-0.15	&	-0.03	&	0.02	&	-0.02	&	-0.06	&	-0.07	&	-0.07\\ 
2M08513577+1153347	&	0.04	&	-0.13	&	0.24	&	0.09	&	0.12	&	0.05	&	0.24	&	0.15	&	-0.06	&	0.01	&	0.02	&	0.14	&	-0.04	&	0.02	&	0.04\\ 
2M08513938+1151456	&	0.06	&	-0.18	&	0.30	&	0.12	&	0.22	&	0.08	&	0.36	&	0.19	&	0.00	&	0.05	&	0.07	&	0.21	&	-0.05	&	0.03	&	0.03\\ 
2M08514234+1150076	&	0.12	&	-0.17	&	0.44	&	0.21	&	0.31	&	0.26	&	0.32	&	0.23	&	-0.03	&	0.15	&	0.13	&	-0.06	&	0.00	&	0.05	&	0.08\\ 
2M08514388+1156425	&	0.07	&	-0.18	&	0.30	&	0.11	&	0.29	&	0.13	&	0.41	&	0.27	&	-0.02	&	0.07	&	0.06	&	0.14	&	-0.04	&	0.07	&	0.02\\ 
2M08514507+1147459	&	0.07	&	-0.14	&	0.29	&	0.11	&	0.33	&	0.10	&	0.34	&	0.21	&	-0.01	&	0.07	&	0.04	&	0.24	&	-0.00	&	0.03	&	0.04\\ 
2M08514883+1156511	&	0.04	&	-0.15	&	0.18	&	0.14	&	0.19	&	0.04	&	0.25	&	0.18	&	-0.10	&	0.03	&	-0.04	&	0.14	&	-0.10	&	-0.02	&	0.04\\ 
2M08515611+1150147	&	0.15	&	-0.17	&	0.09	&	0.15	&	0.27	&	0.18	&	0.42	&	0.28	&	0.02	&	0.19	&	0.14	&	0.37	&	0.05	&	0.08	&	0.10\\ 
2M08515952+1155049	&	0.04	&	-0.17	&	0.31	&	0.07	&	0.26	&	0.08	&	0.39	&	0.23	&	-0.06	&	0.04	&	-0.01	&	0.16	&	-0.05	&	0.02	&	-0.00\\ 
2M08521097+1131491	&	0.10	&	-0.16	&	0.35	&	0.17	&	0.32	&	0.17	&	0.43	&	0.26	&	0.02	&	0.13	&	0.14	&	0.22	&	0.01	&	0.08	&	0.05\\ 
2M08521656+1119380	&	0.03	&	-0.19	&	0.27	&	0.12	&	0.37	&	0.14	&	0.43	&	0.22	&	-0.03	&	0.09	&	0.09	&	0.04	&	-0.07	&	0.04	&	-0.02\\ 
2M08521856+1144263	&	0.06	&	-0.23	&	0.39	&	0.12	&	0.39	&	0.14	&	0.35	&	0.23	&	-0.05	&	0.05	&	0.05	&	0.34	&	-0.04	&	0.07	&	0.05\\ 
2M08522636+1141277	&	0.05	&	-0.19	&	0.33	&	0.20	&	0.17	&	0.10	&	0.25	&	0.20	&	-0.04	&	0.04	&	0.10	&	0.14	&	-0.04	&	-0.04	&	0.06\\ 
2M08525625+1148539	&	0.16	&	-0.21	&	0.31	&	0.31	&	0.29	&	0.21	&	0.43	&	0.31	&	0.07	&	0.15	&	0.23	&	0.21	&	0.08	&	0.12	&	0.16\\ 
2M08534672+1123307	&	0.11	&	0.03	&	0.11	&	0.05	&	0.31	&	0.13	&	0.38	&	0.29	&	0.05	&	0.09	&	0.10	&	0.28	&	0.04	&	0.04	&	0.12\\ 
2M08493465+1151256	&	-1.14	&	-0.72	&	0.68	&	-1.00	&	-2.50	&	-0.98	&	-0.63	&	-1.54	&	-0.72	&	-0.92	&	-1.43	&	-1.63	&	-1.90	&	-0.10	&	-0.97\\ 
2M08505816+1152223	&	0.08	&	-0.05	&	-0.01	&	0.03	&	0.26	&	0.09	&	0.28	&	0.20	&	-0.02	&	0.11	&	0.05	&	0.18	&	-0.01	&	0.05	&	0.09\\ 
2M08510723+1153019	&	-1.71	&	-0.69	&	-0.26	&	-0.15	&	-1.36	&	-2.21	&	-1.77	&	-2.21	&	-1.23	&	-1.53	&	-2.21	&	-0.84	&	-0.89	&	-1.21	&	-1.36\\ 
2M08510839+1147121	&	0.07	&	-0.15	&	-0.08	&	0.26	&	0.16	&	0.09	&	0.27	&	0.21	&	-0.01	&	0.06	&	-0.03	&	0.30	&	0.01	&	0.04	&	0.06\\ 
2M08522003+1127362	&	0.09	&	-0.10	&	-0.07	&	0.27	&	0.28	&	0.13	&	0.30	&	0.22	&	-0.03	&	0.09	&	0.07	&	0.15	&	-0.01	&	0.04	&	0.09\\ 
\textbf{Subgiant} \\
2M08504994+1149127	&	-0.05	&	-0.02	&	0.20	&	-0.05	&	-0.07	&	-0.03	&	0.04	&	0.05	&	-0.09	&	-0.06	&	-0.10	&	-0.04	&	0.01	&	-0.11	&	-0.04\\ 
2M08510325+1145473	&	0.01	&	0.01	&	-0.02	&	0.07	&	-0.76	&	-0.25	&	-0.03	&	-0.02	&	-0.07	&	0.05	&	-0.04	&	-0.52	&	-2.35	&	-0.06	&	-0.01\\ 
2M08511564+1150561	&	0.04	&	-0.11	&	0.10	&	0.15	&	0.16	&	0.02	&	0.19	&	0.11	&	-0.00	&	0.00	&	-0.00	&	0.14	&	-0.07	&	0.02	&	0.07\\ 
2M08511670+1145293	&	0.01	&	-0.02	&	-0.00	&	0.04	&	0.24	&	-0.11	&	0.00	&	0.05	&	-0.05	&	0.02	&	0.03	&	-0.40	&	0.05	&	-0.02	&	-0.00\\ 
2M08512122+1145526	&	-0.05	&	-0.02	&	-0.05	&	0.02	&	-0.55	&	-0.24	&	-0.06	&	-0.11	&	-0.07	&	-0.06	&	-0.26	&	-0.36	&	-2.34	&	-0.10	&	-0.01\\ 
2M08512879+1151599	&	0.00	&	-0.04	&	-0.01	&	-0.02	&	0.29	&	-0.11	&	0.04	&	0.08	&	-0.19	&	-0.01	&	0.09	&	-0.02	&	-0.02	&	-0.13	&	-0.01\\ 
2M08512935+1145275	&	0.01	&	-0.14	&	0.16	&	0.08	&	0.24	&	0.00	&	0.12	&	0.12	&	-0.06	&	-0.01	&	-0.11	&	-0.14	&	-0.03	&	-0.02	&	0.01\\ 
2M08513540+1157564	&	0.01	&	-0.26	&	0.03	&	0.17	&	0.17	&	-0.08	&	0.11	&	0.03	&	-0.01	&	-0.06	&	-0.21	&	0.03	&	-0.07	&	-0.04	&	0.03\\ 
2M08513862+1220141	&	0.02	&	0.03	&	-0.03	&	0.09	&	0.17	&	0.01	&	0.13	&	0.08	&	-0.07	&	-0.04	&	-0.12	&	0.21	&	-0.03	&	0.01	&	0.06\\ 
2M08514401+1146245	&	0.07	&	0.01	&	0.15	&	-0.07	&	0.35	&	0.10	&	0.13	&	0.10	&	-0.03	&	-0.09	&	0.18	&	-0.39	&	0.01	&	0.05	&	0.05\\ 
2M08514474+1146460	&	-0.01	&	-0.17	&	0.14	&	0.11	&	0.19	&	-0.03	&	0.08	&	0.03	&	-0.09	&	-0.05	&	-0.12	&	0.09	&	-0.09	&	-0.02	&	0.03\\ 
2M08514994+1149311	&	-0.00	&	0.01	&	-0.01	&	0.12	&	-2.38	&	0.02	&	0.15	&	0.17	&	0.01	&	0.06	&	0.10	&	0.13	&	-2.25	&	-0.06	&	0.02\\ 
2M08515335+1148208	&	-0.06	&	-0.06	&	-0.07	&	0.05	&	-0.95	&	-0.04	&	0.10	&	0.07	&	-0.01	&	-0.01	&	0.06	&	0.04	&	-2.17	&	-0.12	&	-0.02\\ 
2M08521134+1145380	&	0.08	&	-0.07	&	0.29	&	0.03	&	0.39	&	0.10	&	0.07	&	0.12	&	-0.03	&	0.04	&	0.06	&	-0.29	&	-0.02	&	0.02	&	0.06\\ 
2M08503667+1148553	&	-0.07	&	-0.04	&	-0.06	&	-0.09	&	-1.34	&	-0.09	&	0.12	&	-0.02	&	0.05	&	0.01	&	0.03	&	-0.03	&	-2.38	&	-0.20	&	-0.07\\ 
2M08505569+1152146	&	0.01	&	0.02	&	-0.08	&	0.00	&	0.37	&	0.02	&	0.15	&	0.15	&	-0.03	&	0.01	&	-0.06	&	0.14	&	-0.26	&	-0.11	&	0.01\\ 
2M08510106+1150108	&	0.08	&	0.04	&	-0.08	&	0.34	&	0.15	&	0.08	&	0.16	&	0.18	&	-0.03	&	0.09	&	0.17	&	-0.04	&	0.08	&	0.04	&	0.06\\ 
2M08510951+1141449	&	0.15	&	0.07	&	-0.04	&	0.25	&	0.23	&	0.13	&	0.23	&	0.27	&	0.01	&	0.16	&	0.22	&	-0.15	&	0.18	&	0.15	&	0.15\\ 
2M08511877+1151186	&	0.09	&	-0.08	&	-0.01	&	-0.06	&	0.17	&	0.08	&	0.28	&	0.15	&	0.04	&	0.08	&	0.04	&	0.21	&	-0.02	&	0.04	&	0.11\\ 
2M08515567+1217573	&	0.04	&	-0.07	&	-0.07	&	0.14	&	-2.20	&	0.03	&	0.20	&	0.14	&	-0.10	&	0.05	&	-0.04	&	0.09	&	0.01	&	-0.01	&	0.06\\ 
\textbf{turnoff} \\
2M08503392+1146272	&	0.01	&	0.01	&	...	&	...	&	...	&	-0.00	&	0.14	&	0.10	&	-0.05	&	-0.04	&	-0.31	&	0.01	&	...	&	-0.04	&	0.05\\ 
2M08504079+1147462	&	-0.05	&	-0.03	&	...	&	...	&	...	&	-0.04	&	0.07	&	0.13	&	-0.09	&	-0.03	&	0.22	&	-0.31	&	...	&	-0.18	&	-0.02\\ 
2M08505177+1200247	&	-0.03	&	-0.03	&	...	&	...	&	...	&	-0.22	&	-0.04	&	-0.10	&	-0.19	&	-0.09	&	-0.47	&	0.05	&	...	&	-0.09	&	-0.01\\ 
2M08505702+1159158	&	-0.08	&	-0.07	&	...	&	...	&	...	&	-0.26	&	-0.03	&	-0.13	&	0.03	&	-0.03	&	-0.06	&	-0.10	&	...	&	-0.14	&	-0.02\\ 
2M08505762+1155147	&	-0.02	&	-0.02	&	...	&	...	&	...	&	-0.08	&	0.04	&	0.06	&	-0.01	&	-0.05	&	-0.36	&	-0.10	&	...	&	-0.09	&	-0.00\\ 
2M08505903+1148576	&	-0.03	&	-0.02	&	...	&	...	&	...	&	-0.26	&	-0.10	&	-0.02	&	-0.03	&	-0.07	&	-0.53	&	0.03	&	...	&	-0.11	&	-0.05\\ 
2M08505973+1139524	&	-0.03	&	-0.02	&	...	&	...	&	...	&	-0.22	&	-0.08	&	0.01	&	-0.15	&	-0.02	&	-0.32	&	-0.35	&	...	&	-0.11	&	0.00\\ 
2M08510969+1159096	&	-0.11	&	-0.12	&	...	&	...	&	...	&	-0.34	&	-0.15	&	-0.15	&	-0.10	&	-0.09	&	-0.20	&	-0.28	&	...	&	-0.19	&	-0.12\\ 
2M08511576+1152587	&	-0.02	&	-0.00	&	...	&	...	&	...	&	-0.18	&	0.13	&	-0.04	&	0.06	&	0.02	&	0.01	&	-0.04	&	...	&	-0.05	&	0.04\\ 
2M08512240+1151291	&	-0.12	&	-0.11	&	...	&	...	&	...	&	-0.28	&	-0.18	&	-0.17	&	-0.06	&	-0.16	&	-0.22	&	0.03	&	...	&	-0.16	&	-0.05\\ 
2M08513710+1154599	&	-0.01	&	-0.07	&	...	&	...	&	...	&	-0.15	&	-0.03	&	0.05	&	-0.10	&	0.11	&	-0.11	&	-0.15	&	...	&	-0.10	&	0.06\\ 
2M08513806+1201243	&	-0.01	&	-0.04	&	...	&	...	&	...	&	-0.25	&	-0.10	&	-0.06	&	-0.02	&	0.11	&	-0.28	&	-0.05	&	...	&	-0.12	&	0.04\\ 
2M08514122+1154290	&	-0.04	&	-0.03	&	...	&	...	&	...	&	-0.22	&	-0.01	&	-0.06	&	-0.01	&	-0.05	&	-0.49	&	0.04	&	...	&	-0.12	&	-0.00\\ 
2M08514475+1145012	&	-0.10	&	-0.08	&	...	&	...	&	...	&	-0.32	&	-0.19	&	-0.11	&	-0.04	&	-0.14	&	-0.14	&	0.26	&	...	&	-0.13	&	-0.06\\ 
2M08520741+1150221	&	-0.00	&	-0.01	&	...	&	...	&	...	&	-0.13	&	0.05	&	0.02	&	-0.12	&	0.02	&	-0.36	&	-0.10	&	...	&	-0.03	&	0.02\\ 
\textbf{main-sequence} \\
2M08502805+1154505	&	0.02	&	-0.00	&	...	&	...	&	...	&	-0.10	&	0.02	&	-0.08	&	-0.03	&	0.06	&	0.12	&	-0.12	&	...	&	-0.06	&	-0.03\\ 
2M08511229+1154230	&	0.04	&	0.01	&	...	&	...	&	...	&	-0.08	&	0.01	&	0.03	&	0.06	&	0.06	&	-0.00	&	0.06	&	...	&	-0.04	&	0.05\\ 
2M08512314+1154049	&	0.00	&	-0.04	&	...	&	...	&	...	&	-0.17	&	-0.08	&	-0.01	&	-0.01	&	0.05	&	-0.49	&	0.04	&	...	&	-0.12	&	0.01\\ 
2M08512604+1149555	&	-0.02	&	-0.13	&	...	&	...	&	...	&	-0.05	&	0.15	&	-0.08	&	-0.01	&	0.03	&	0.03	&	0.02	&	...	&	-0.12	&	-0.02\\ 
2M08512996+1151090	&	-0.01	&	-0.03	&	...	&	...	&	...	&	-0.14	&	-0.04	&	-0.00	&	0.01	&	0.04	&	-0.07	&	0.03	&	...	&	-0.05	&	0.03\\ 
2M08513119+1153179	&	-0.03	&	-0.03	&	...	&	...	&	...	&	-0.19	&	-0.10	&	-0.04	&	-0.07	&	-0.06	&	-0.19	&	0.04	&	...	&	-0.07	&	-0.01\\ 
2M08513701+1136516	&	0.03	&	0.06	&	...	&	...	&	...	&	0.03	&	0.02	&	0.04	&	-0.01	&	0.01	&	0.12	&	0.15	&	... &	0.03	&	0.07\\ 
2M08514189+1149376	&	0.06	&	0.02	&	...	&	...	&	...	&	-0.10	&	-0.09	&	0.12	&	-0.05	&	0.14	&	-0.11	&	0.07	&	...	&	0.02	&	0.11\\ 
2M08514742+1147096	&	0.05	&	-0.03	&	...	&	...	&	...	&	0.00	&	0.03	&	0.01	&	-0.00	&	0.02	&	0.01	&	0.08	&	...	&	0.03	&	0.03\\ 
2M08521649+1147382	&	0.02	&	-0.04	&	...	&	...	&	...	&	-0.15	&	-0.06	&	0.07	&	-0.03	&	0.05	&	-0.33	&	0.13	&	...	&	-0.01	&	0.07\\ 
2M08505439+1156290	&	-0.04	&	-0.02	&	...	&	...	&	...	&	-0.21	&	-0.01	&	-0.08	&	-0.11	&	-0.08	&	-0.11	&	-0.04	&	...	&	-0.10	&	0.04\\ 
2M08510076+1153115	&	0.02	&	-0.02	&	...	&	...	&	...	&	-0.13	&	0.10	&	0.04	&	-0.02	&	-0.03	&	-0.08	&	0.11	&	...	&	-0.05	&	0.05\\ 
2M08511176+1150018	&	0.03	&	-0.01	&	...	&	...	&	...	&	-0.32	&	-0.10	&	0.04	&	-0.09	&	0.12	&	-0.07	&	0.10	&	...	&	-0.05	&	0.04\\ 
2M08512080+1145024	&	-0.03	&	-0.03	&	...	&	...	&	...	&	-0.21	&	-0.12	&	-0.02	&	-0.24	&	0.03	&	-0.20	&	-0.84	&	...	&	-0.13	&	-0.04\\ 
2M08512742+1153265	&	-0.04	&	-0.03	&	...	&	...	&	...	&	-0.15	&	0.01	&	-0.03	&	-0.01	&	-0.01	&	-0.04	&	-0.02	&	...	&	-0.07	&	0.02\\ 
2M08512788+1155409	&	-0.03	&	-0.02	&	...	&	...	&	...	&	-0.20	&	-0.01	&	-0.11	&	0.02	&	0.01	&	-0.09	&	0.04	&	...	&	-0.07	&	0.09\\ 
2M08513012+1143498	&	-0.07	&	-0.06	&	...	&	...	&	...	&	-0.28	&	-0.19	&	-0.06	&	-0.25	&	-0.05	&	-0.56	&	-0.75	&	...	&	-0.17	&	-0.07\\ 
2M08513455+1149068	&	0.06	&	0.02	&	...	&	...	&	...	&	-0.12	&	0.01	&	0.09	&	-0.01	&	0.13	&	0.05	&	0.06	&	...	&	-0.01	&	0.09\\ 
2M08521868+1143246	&	-0.08	&	-0.05	&	...	&	...	&	...	&	-0.35	&	-0.14	&	-0.07	&	-0.08	&	-0.10	&	-0.17	&	0.21	&	...	&	-0.21	&	-0.10\\ 
2M08512643+1143506	&	-0.21	&	-0.67	&	...	&	...	&	...	&	-0.16	&	-1.56	&	0.26	&	0.13	&	0.58	&	-0.66	&	-2.11	&	...	&	0.54	&	0.04\\ 
2M08513259+1148520	&	-0.18	&	-0.15	&	...	&	...	&	...	&	-0.17	&	-0.46	&	0.02	&	-0.60	&	-0.61	&	-0.55	&	-0.61	&	...	&	-0.19	&	-0.15\\ 
\hline
Excluded Sample due\\
to low SNR ($<$) 100\\
\textbf{Subgiant} \\
2M08503438+1139566	&	0.10	&	0.07	&	0.06	&	0.09	&	0.72	&	0.03	&	0.24	&	0.25	&	-0.05	&	0.20	&	0.16	&	-0.08	&	-0.25	&	-0.08	&	0.08\\ 
2M08504198+1136525	&	0.11	&	0.03	&	0.07	&	0.03	&	0.33	&	0.08	&	0.11	&	0.34	&	0.02	&	0.26	&	0.24	&	-0.04	&	0.00	&	-0.01	&	0.12\\ 
2M08510811+1201065	&	-0.07	&	-0.13	&	-0.12	&	-0.32	&	...	&	-0.13	&	0.36	&	0.03	&	-0.09	&	0.18	&	0.29	&	-0.26	&	...	&	-0.18	&	-0.05\\ 
2M08511826+1150196	&	-0.17	&	-0.21	&	-0.20	&	-0.39	&	...	&	-0.29	&	0.07	&	-0.12	&	-0.15	&	-0.03	&	-0.12	&	-0.24	&	...	&	-0.30	&	-0.21\\ 
2M08520356+1141238	&	-0.04	&	-0.02	&	-0.05	&	0.01	&	-0.56	&	0.04	&	0.08	&	0.11	&	0.05	&	-0.01	&	0.20	&	0.09	&	-0.22	&	-0.27	&	0.01\\ 
\textbf{main-sequence} \\
2M08502833+1142097	&	0.01	&	-0.02	&	...	&	...	&	...	&	-0.20	&	-0.05	&	0.16	&	0.05	&	0.05	&	0.27	&	-0.05	&	...	&	0.05	&	0.09\\ 
2M08503788+1252295	&	0.07	&	0.04	&	...	&	...	&	...	&	-0.18	&	-0.05	&	0.08	&	-0.03	&	0.12	&	-0.06	&	0.07	&	...	&	-0.01	&	0.14\\ 
2M08505334+1143399	&	-0.10	&	-0.13	&	...	&	...	&	...	&	-0.60	&	-0.11	&	-0.07	&	-0.36	&	0.04	&	-0.41	&	-0.50	&	...	&	-0.18	&	-0.05\\ 
2M08505923+1146129	&	-0.10	&	-0.11	&	...	&	...	&	...	&	-0.43	&	-0.47	&	-0.18	&	-0.18	&	-0.07	&	-0.14	&	-0.85	&	...	&	-0.10	&	-0.04\\ 
2M08512386+1138521	&	-0.04	&	-0.07	&	...	&	...	&	...	&	-0.25	&	-0.14	&	-0.09	&	-0.10	&	-0.10	&	-0.47	&	0.07	&	...	&	-0.10	&	0.02\\ 
2M08513215+1136126	&	-0.05	&	-0.03	&	...	&	...	&	...	&	-0.20	&	-0.16	&	-0.14	&	-0.06	&	-0.32	&	0.48	&	-2.50	&	...	&	0.04	&	0.01\\ 
2M08513444+1137574	&	0.02	&	0.02	&	...	&	...	&	...	&	-0.28	&	-0.10	&	0.01	&	-0.24	&	0.05	&	0.35	&	0.75	&	...	&	-0.04	&	-0.02\\ 
2M08514375+1145148	&	-0.04	&	-0.08	&	...	&	...	&	...	&	-0.22	&	-0.17	&	-0.02	&	0.06	&	-0.05	&	0.21	&	0.41	&	...	&	0.00	&	0.00\\ 
2M08514465+1141510	&	-0.02	&	0.02	&	...	&	...	&	...	&	-0.19	&	-0.19	&	0.03	&	-0.07	&	0.03	&	0.17	&	0.43	&	...	&	-0.16	&	-0.00\\ 
2M08515290+1146358	&	0.12	&	0.02	&	...	&	...	&	...	&	-0.15	&	-0.11	&	0.25	&	-0.05	&	0.22	&	0.23	&	0.38	&	...	&	0.02	&	0.07\\ 
2M08521664+1142300	&	-0.15	&	-0.10	&	...	&	...	&	...	&	-0.31	&	-0.16	&	-0.13	&	0.22	&	-0.16	&	0.04	&	0.29	&	...	&	-0.29	&	-0.09\\ 
2M08504511+1136023	&	0.05	&	-0.08	&	...	&	...	&	...	&	-0.34	&	-0.34	&	0.12	&	0.04	&	0.03	&	-0.13	&	-0.27	&	...	&	-0.08	&	-0.15\\ 
2M08510131+1141587	&	-0.17	&	-0.19	&	...	&	...	&	...	&	-0.42	&	-0.23	&	-0.13	&	-0.39	&	-0.11	&	-0.43	&	-0.36	&	...	&	-0.23	&	-0.10\\ 
2M08510156+1147501	&	-0.09	&	-0.06	&	...	&	...	&	...	&	-0.42	&	-0.08	&	-0.25	&	-0.10	&	-0.06	&	-0.07	&	-0.54	&	...	&	-0.17	&	-0.00\\ 
2M08511229+1146212	&	-0.05	&	-0.02	&	...	&	...	&	...	&	-0.29	&	-0.08	&	-0.12	&	-0.14	&	0.05	&	-0.03	&	-0.40	&	...	&	-0.04	&	0.01\\ 
2M08511810+1142547	&	-0.02	&	0.01	&	...	&	...	&	...	&	-0.22	&	-0.09	&	-0.06	&	-0.14	&	-0.03	&	-0.32	&	-0.82	&	...	&	-0.07	&	-0.01\\ 
2M08512033+1145523	&	-0.03	&	-0.01	&	...	&	...	&	...	&	-0.21	&	-0.06	&	-0.01	&	-0.22	&	0.01	&	0.03	&	-0.32	&	...	&	-0.05	&	-0.03\\ 
2M08512176+1144050	&	-0.05	&	-0.06	&	...	&	...	&	...	&	-0.48	&	-0.12	&	-0.10	&	-0.10	&	0.09	&	-0.07	&	-0.10	&	...	&	-0.10	&	-0.09\\ 
2M08512467+1143061	&	0.03	&	0.01	&	...	&	...	&	...	&	-0.47	&	-0.05	&	0.13	&	-0.34	&	0.11	&	-0.38	&	0.21	&	...	&	-0.16	&	-0.07\\ 
2M08513424+1145535	&	0.06	&	0.03	&	...	&	...	&	...	&	-0.03	&	0.10	&	0.04	&	-0.01	&	0.13	&	0.08	&	0.14	&	...	&	0.02	&	0.09 \\
\tablewidth{0pt}	
\enddata
\end{deluxetable}
\end{longrotatetable}

\end{document}